\documentclass[letterpaper,11pt]{article}
\usepackage{jheppub}
\usepackage{amsmath,latexsym}
\usepackage{booktabs}
\usepackage{graphicx}
\usepackage{siunitx}
\usepackage{setspace}
\usepackage{amssymb}
\usepackage{amstext}
\usepackage{amsfonts}
\usepackage{bbold}
\usepackage{slashed}
\usepackage{tikz-feynman}
\usepackage{xcolor}
\usepackage{subcaption}
\usepackage{physics}
\newcommand{\beq}{\begin{equation}}
\newcommand{\eeq}{\end{equation}}
\renewcommand{\l}{\left}
\renewcommand{\r}{\right}

\newcommand{\bea}{\begin{eqnarray}}
\newcommand{\eea}{\end{eqnarray}}

\newcommand{\nn}{\nonumber}

\newcommand{\be}{\begin{eqnarray}}
\newcommand{\ee}{\end{eqnarray}}
\newcommand{\cP}{{\cal P}}

\definecolor{SGreen}{rgb}{0.0547,0.613,0.328}
\definecolor{NodeBlue}{rgb}{0.0547,0.148,0.578}

\newcommand{\dbar}{d\hspace*{-0.08em}\bar{}\hspace*{0.1em}}
\tikzfeynmanset{warn luatex=false}
\begin{document}

\title{Relations Between Anomalous Dimensions in the Regge Limit} 

\author{Ira Z. Rothstein and}
\author{Michael Saavedra}
\affiliation{Department of Physics, Carnegie Mellon University, Pittsburgh, PA 15213, USA}

\abstract{

We extend the   recent formalism  developed for computing rapidity anomalous dimension  of form factors using unitarity to the problem of high-energy near forward scattering.  By combining the factorization of $2\rightarrow 2$ scattering in the effective field theory (EFT) for Glauber operators with definite signature amplitudes, we derive an expression that relates anomalous dimensions (including Regge trajectories) to cut amplitudes, leading to significant computational simplifications.  We demonstrate this explicitly by computing the one and two-loop Regge trajectories. Our formalism can also be used to bootstrap   anomalous dimensions of operators not related by symmetries.   As an example, we show that the full  anomalous dimensions (including both the Regge pole and cut pieces) of the  two Glauber operator anti-symmetric octet operator, can be determined from the anomalous dimension of the  single Glauber exchange operator. Many other such relations exist between other color channels at each order in $\alpha$.}

\maketitle

\section{Introduction}

Despite the remarkable progress made  in our understanding  of resummed perturbative field theory, it is fair to
say that when it comes to the near forward scattering  (Regge) limit, $s\gg t$  there are still many open questions. 
Gribov's original approach \cite{Gribov:1967vfb} to  the problem has  led to a  number of perspectives including the classic work of Balitsky, Fadin, Kuraev and Lipatov \cite{Fadin:1975cb}, Lipatov’s effective action \cite{Lipatov:1995pn}, and more modern approaches in terms of Wilson lines\cite{Balitsky:1995ub,Caron-Huot:2013fea,Caron-Huot:2017fxr, Vernazza:2018gyb, Falcioni:2021buo}.

What complicates
the perturbative series in the Regge limit is the existence of large logs  of the ratio $s/t$ that appear at each order in perturbation theory.
The resummation of large logs  is not an exotic phenomena as logs of the ratios of invariant masses
 are summed via 
a canonical renormalization 
group analysis which follows from factorization of mass scales, or equivalently decoupling.
However,   Regge logs grow as a ratio of
rapidities and are not summable in this way. 

 The resummations of rapidity logs has a long history.
In the context of hard scattering, the Collins-Soper equation \cite{Collins:1981uk} 
resums logs in hard scattering transverse momentum distributions, while  the BFKL \cite{Kuraev:1977fs,Balitsky:1978ic}  and its generalization the  BJIMWLK equation \cite{Jalilian-Marian:1997qno,Iancu:2000hn}  resum  rapidity logs\footnote{ We will use the acronym RRG to refer to the rapidity renormalization group equation.} in  near forward scattering  processes.  
A  universal  formalism 
which allows for the  resummation of rapidity logs for both hard scattering and in the  Regge limit was developed in \cite{Chiu:2011qc,Chiu2012ATheory}.
The universality of this approach stems from  fact that it is based upon  an operator formalism within the confines of an effective field theory, which in the case 
 of hard scattering, corresponds to SCET \cite{Bauer:2000ew,Bauer:2001cu,Bauer:2001yt} and  its' generalization in the near forward  scattering limit (GSCET) \cite{Rothstein:2016bsq} \footnote{The Glauber mode breaks factorization in the case
 of hard scattering. In all SCET proofs to date \cite{Bauer:2002nz}  utilized the version of SCET which did not include Glaubers\cite{Bauer:2000ew}, it was simply assumed (given the proofs of Collins, Soper and Sterman (\cite{Collins:1988ig}) in the full theory) that the Glauber mode cancelled in the relevant observables.} .
The EFT approach to forward scattering systematizes the problem in the sense that the resummation of rapidity logs  is reduced
to finding the  anomalous dimensions for a  set of well defined gauge invariant operators in the effective theory \footnote{The EFT approach is related to but quite distinct from the Reggeon field theory approach. For instance, Reggeon exchange should not be equated with Glauber exchange. For a discussion of these distinctions see \cite{Gao2024}.}.

What makes the Regge problem more complex than the case of hard scattering is that at each order of perturbation theory, there are new
operators, broken up into irreducible representations of  color of various dimensions, that appear at the same (leading) order  in $t/s$ which leads to a complex mixing matrix for non-local operators.
However, it  has been known for a long time that  the perturbative series in the Regge limit has a very rich structure  that
can serve to greatly simplify the system. 
This structure  is manifested in the results found  \cite{Kuraev:1976ge} for  the gluon near forward amplitude 
in the anti-symmetric octet ($8_A$) channel.
In particular, it was shown that at NLL the  glue-glue scattering amplitude in this channel takes the form
\label{Fadin}
\begin{eqnarray}
    \mathcal{M}^{8_A}_{2\rightarrow 2} &=& [g_sT^c_{aa^\prime} C_g(p_a,p_{a^\prime})]
    \frac{s}{t}
   \l[ \l(\frac{s + i\epsilon}{-t}\r)^{\alpha(t)} 
    +\l(\frac{-s + i\epsilon}{-t}\r)^{\alpha(t)} \r]
  [ g_aT^c_{bb^\prime} C_g(p_b,p_{b^\prime})]\nn \\
     &=& [g_sT^c_{aa^\prime} C_g(p_a,p_{a^\prime})]
    \frac{s}{t}
   \l[ \l(\frac{s + i\epsilon}{-t}\r)^{\alpha(t)} (
   1 +e^{-i \pi \alpha(t)}) \r]
  [ g_aT^c_{bb^\prime} C_g(p_b,p_{b^\prime})]
    \label{Planar Factorization}
\end{eqnarray}
with $C_g(t)$ being the so-called "impact factor" 
and $\alpha(t)$ is the gluon Regge trajectory.
Note that the result is anti-symmetric under crossing  in the kinematic variable and Bose symmetry follows since the color
factor is crossing odd.
  Similar results  holds for the case of the  non-crossing symmetric quark scattering  amplitude\cite{Kuraev:1976ge}, albeit with different impact factors $C_q$ and representations of the color generators.
For details on the structure at two loops and the breakdown of the result beyond NLL see \cite{Duca_2001}.
This result  is valid to next to leading log (NLL) and is exact in the planar limit. It is fair to say that this result seems unexpected as it implies several remarkable facts. The power law in $s/t$ has the form of a solution to a simple differential equation which arises when running local operators, whereas non-local operators obey
 integro-differential equations whose solutions are in general not simple power laws. Such power laws correspond to ``Regge poles'' as opposed to ``Regge cuts" that arise in the complex angular momentum plane (see  e.g. \cite{Kovchegov:2012mbw}). Furthermore, there seems to be only one quantity associated with an ``anomalous dimension", $\alpha(t)$ ,while we would expect at least two such anomalous dimensions at two loops, and more beyond.
 Other hints of underlying structure were pointed out in \cite{Del_Duca_2018} where it was shown   that the constants
appearing in the one-loop quark and gluon impact factors involve the  finite part of the two-loop Regge trajectory,
though the impact factors are polluted by additional constants that are not related to the Regge trajectory.
Intuitively this iterative structure would seem to be a consequence of unitarity as forward scattering is a semi-classical process which leads to the exponentiation of the classical (shock wave) action.
On the other hand the anomalous dimensions and the associated logs are  a quantum effect so there is no apriori reason to expect any natural relation between the iterative (phase) structure and the aforementioned anomalous dimensions relation. The are other interesting relations between the lower loop results and higher order contributions to the cusp anomalous
dimensions  \cite{Korchemskaya:1996je,Korchemskaya:1994qp,DelDuca:2011wkl,DelDuca:2011ae}, which will not be relevant to our discussion.

Recently it was shown \cite{Moult:2022lfy} that the iterative  structure is elegantly exposed within the EFT formalism \cite{Rothstein:2016bsq}.
In particular, the authors used the  EFT approach  to show that,  given the form of the amplitude (\ref{Fadin}), 
the  finite part of the two-loop Regge trajectory must be encoded  in the order $\epsilon$ piece of a one loop correction (with no contamination)  while
the divergent pieces arise from a simplified (in a manner to made clear below) two loop calculation which, at the technical level, is 
effectively one loop.  Their results also shows that in the EFT one reduces the calculation of the two loop Regge trajectory to
three diagrams. Finally,  the authors  were able to  derive the maximally matter dependent contributions to the
Regge trajectory to all loop orders.

In this paper, we will generalize the result in  \cite{Moult:2022lfy} in several ways. Firstly due to the reliance of  \cite{Moult:2022lfy}
on the use of the form of the amplitude (\ref{Fadin}) their results can only  used  in  the $8_A$ channel and only to NLL, away from the planar limit, whereas our relations will be valid in all channels and to all orders.  In allowing for  more general processes we will also uncover additional relations among various anomalous dimensions
which will contribute to Regge cuts. Furthermore, our results show that  simplifications  in calculating the  Regge trajectory, in the case of anti-symmetric octet  are universal and, moreover, the Regge trajectory as well as other anomalous dimensions, can be calculated directly via cut diagrams in the EFT. That Regge logs can be extracted by cut diagrams in the full theory was shown  
in \cite{Fadin:1995xg}, though working in the EFT systematizes the methodology, and streamlines the calculations, especially at higher orders.

We will begin by reviewing the EFT for hard scattering   SCET \cite{Bauer:2000ew,Bauer:2001cu,Bauer:2001yt},  generalized to allow for forward scattering \cite{Rothstein:2016bsq}.
The reader looking for more details regarding the rudiments of SCET may consult \cite{SCETreview,Becher:2014oda}.

\section{Conventions}
We use the mostly minus metric $g_{\mu \nu}=Diag(1,-1,-1,-1)$ 
and define the usual $\overline{\text{MS}}$ factor
\begin{equation}
\bar{\mu}^{2\epsilon} = \mu^{2\epsilon} (4\pi)^{-\epsilon}e^{\epsilon \gamma_E},
\end{equation}
where we use $d=4-2 \epsilon$ and 
\beq
[d^dk]= \frac{d^dk}{(2 \pi)^d}.
\eeq
For transverse momentum integration, we will use the notation $d^\prime=2-2\epsilon$.

For the light cone coordinates we define two null vectors $n/\bar n^\mu=(1,0,0,\pm 1)$
and decompose four vectors as follows
\beq
p^\mu= n \cdot p \frac{\bar n^\mu}{2}+ \bar n \cdot p \frac{n^\mu}{2}+p_\perp^\mu \equiv (n\cdot p,\, \bar n\cdot p,\,p_\perp^\mu),
\eeq
such that 
\begin{equation}
p^-=n \cdot p, ~~ p^+=\bar n \cdot p .
\end{equation}

The phase space on shell delta function will be written as
\begin{equation}
\delta_+(p^2 - m^2) = \delta(p^2 - m^2) \bar \theta(\bar{n}\cdotp p + n\cdotp p),
\end{equation}
and $\bar \theta(x)\equiv (2 \pi) \theta(x)$. 

We will be focusing on 2-to-2 scattering in the limit of $s>>|t|$, $s>0$, $t<0$, which in the $s$-channel we take to be $(p_1,p_2\rightarrow p_3, p_4)$,  
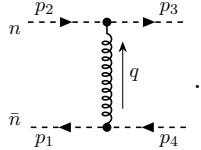
\be
\begin{gathered}
\scalebox{0.7}{
\begin{tikzpicture}
\begin{feynman}
	\node[dot] (f1) at (0,1);
	\node[dot] (f2) at (0,-1);
	\vertex[label =above left: \(p_3\)] (p3) at (1.5, 1);
	\vertex[label =above right: \(p_2\)] (p2) at (-1.5, 1);
	\vertex[label =below right: \(p_1\)] (p1) at (-1.5, -1);
	\vertex[label =below left: \(p_4\)] (p4) at (1.5, -1);
	\vertex [label= \(n\)] at (-1.75, .64);
	\vertex [label= \(\bar{n}\)] at (-1.75, -1.1);
\diagram*{
	(f2)--[gluon, line width = 0.3mm, momentum' = \(q\)](f1),
	(p2)--[charged scalar,  line width = 0.3mm](f1)--[charged scalar,  line width = 0.3mm](p3),
	(p4)--[charged scalar,  line width = 0.3mm](f2)--[charged scalar,  line width = 0.3mm](p1),
};
\end{feynman}
\end{tikzpicture}
}
\end{gathered}.\nonumber
\ee
We work in a frame such that $q = q_\perp$ and 
\begin{equation}
    p_1 = (n\cdotp p_1, \bar{n}\cdotp p_1, \vec{q}_\perp/2),\qquad p_2 = (n\cdotp p_2, \bar{n}\cdotp p_2, -\vec{q}_\perp/2),
\end{equation}
with $p_3 = p_2 + q$ and $p_4 = p_1- q$.  We then have Mandelstams $t = q_\perp^2$ and $s =n\cdotp p_1 \bar{n}\cdotp p_2 + O(t)$.  We will also need the $u$-channel process, which we can take to be $(p_1,\bar{p}_3)\rightarrow \bar{p}_2,p_4)$, which diagrammatically is
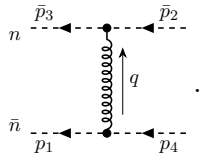
\be
\begin{gathered}
\scalebox{0.7}{
\begin{tikzpicture}
\begin{feynman}
	\node[dot] (f1) at (0,1);
	\node[dot] (f2) at (0,-1);
	\vertex[label =above left: \(\bar{p}_2\)] (p3) at (1.5, 1);
	\vertex[label =above right: \(\bar{p}_3\)] (p2) at (-1.5, 1);
	\vertex[label =below right: \(p_1\)] (p1) at (-1.5, -1);
	\vertex[label =below left: \(p_4\)] (p4) at (1.5, -1);
	\vertex [label= \(n\)] at (-1.75, .64);
	\vertex [label= \(\bar{n}\)] at (-1.75, -1.1);
\diagram*{
	(f2)--[gluon, line width = 0.3mm, momentum' = \(q\)](f1),
	(p3)--[charged scalar,  line width = 0.3mm](f1)--[charged scalar,  line width = 0.3mm](p2),
	(p4)--[charged scalar,  line width = 0.3mm](f2)--[charged scalar,  line width = 0.3mm](p1),
};
\end{feynman}
\end{tikzpicture}
}
\end{gathered}.\nonumber
\ee
In this paper, we take all external states to be quarks, but our main results in section two will apply to any choice of projectiles.
The loop order of objects will appear as a superscript. e.g. $\gamma^{(1)}$ would correspond to a one loop anomalous dimension. However, Glauber loops will not count in this notation.

\section{Glauber SCET}

Our results will all be given in the context of the Glauber variation of SCET designed to study near forward scattering.
The total Lagrangian is written as
\beq
{\cal L}= {\cal L}_n+{\cal L}_{\bar n}+{\cal L}_s+ {\cal L_G}.
\eeq
where  ${\cal L}_n+{\cal L}_{\bar n}+{\cal L}_s$ correspond to the canonical actions \cite{Bauer:2000ew,Bauer:2001cu,Bauer:2001yt} for soft and collinear modes 
while ${\cal L_G}$ accounts for the factorization violating interactions  due to Glauber exchange.

For collinear/anti-collinear  $(n,\bar n)$ scattering via Glauber exchange 
the leading power Glauber operators are
\begin{align}  \label{eq:Onsnb}
     O^{q q}_{ns\bar{n}} &=  {\cal O}_n^{q B}   
     \frac{1}{\cP_\perp^2} {\cal O}_s^{BC} 
     \frac{1}{\cP_\perp^2} 
     {\cal O}_{\bar n}^{q C} \,
    &  O^{g q}_{ns\bar{n}} &=  {\cal O}_n^{g B}   \frac{1}{\cP_\perp^2}  {\cal O}_s^{BC} \frac{1}{\cP_\perp^2}{\cal O}_{\bar n}^{q C} \,,
    \nn\\
        O^{q g}_{ns\bar{n}} &=  {\cal O}_n^{q B} \frac{1}{\cP_\perp^2} {\cal O}_s^{BC} \frac{1}{\cP_\perp^2} {\cal O}_{\bar n}^{g C} \,,
    &  O^{g g}_{ns\bar{n}} &=  {\cal O}_n^{g B} \frac{1}{\cP_\perp^2} {\cal O}_s^{BC} 
    \frac{1}{\cP_\perp^2} {\cal O}_{\bar n}^{g C} \,.
\end{align}
On the left-hand side the subscripts indicate that these operators involve three sectors $\{n,s,\bar{n}\}$, while the first and second superscript determine whether we take a quark or gluon operator in the $n$-collinear or $\bar{n}$-collinear sectors. 
The collinear operators are defined as
\begin{align}  \label{eq:On}
  {\cal O}_n^{q B} &= \overline\chi_{n} T^B \frac{\bar{n}\!\!\!\slash}{2} \: \chi_{n} \,,
  & {\cal O}_n^{g B} &= \biggl[ \frac{i}{2} f^{BCD}  {\cal B}_{n\perp\mu}^C \,
   \frac{\bar{n}}{2}\cdot (\cP\!+\!\cP^\dagger)  {\cal B}_{n\perp}^{D\mu} \biggr]  \,.
\end{align}
These operators are built from the gauge invariant building blocks 
\begin{align}
\chi_n &= W_{n}^\dagger \xi_n \,,     &{\cal B}_{n\perp}^\mu=
      \frac{1}{g}\, \big[ W_n^\dagger i D_{n\perp}^\mu W_n \big] 
\end{align}
where
\beq
 i D_{n\perp}^\mu={\cal P}_\perp^\mu+g A_{n\perp}
\eeq
and $W_n$ is a collinear Wilson line usually written as 
\beq
W_n=\sum_{perms}e^{-\frac{g}{{\cal P}_n} {\bar n} \cdot A_{n \perp}^\mu }.
\eeq
${\cal P}_n$  and  ${\cal P}_\perp$ are the derivative operators which pick out the large light cone and small transverse momenta respectively.

The soft operator is defined as 
\begin{align}  \label{eq:Os1}
    {\cal O}_s^{BC} 
 & ={8\pi\alpha_s  }
    \bigg\{
    {\cal P}_\perp^\mu 
    {\cal S}_n^T
     {\cal S}_{\bar n}
     \cP_{\perp\mu}
    - \cP^\perp_\mu g \widetilde {\cal B}_{S\perp}^{n\mu}  {\cal S}_n^T  {\cal S}_{\bar n}   
    -  {\cal S}_n^T  {\cal S}_{\bar{n}}  g \widetilde {\cal B}_{S\perp}^{\bar{n}\mu} \cP^\perp_{\mu}  
    -  g \widetilde {\cal B}_{S\perp}^{n\mu}  {\cal S}_n^T  {\cal S}_{\bar{n}} g \widetilde {\cal B}_{S\perp\mu}^{\bar{n}}
\nn\\
 &\qquad\qquad
    -\frac{n_\mu {\bar{n}}_\nu}{2} {\cal S}_n^T   ig \widetilde {G}_s^{\mu\nu} {\cal S}_{\bar{n} }
    \bigg\}^{BC} 
    \,.
\end{align}
Note that when no soft gluons are emitted  we have ${\cal O}_s^{BC}=8\pi\alpha_s\delta^{BC} \cP_\perp^2$ and the three sector
operators reduce to tree level Glauber exchange. The one gluon piece of  ${\cal O}_s^{BC}$ is responsible for the Lipatov vertex
whose Feynman rules can be found in \cite{Rothstein:2016bsq}.
The soft gauge invariant operators a in the fundamental representation
\begin{align} \label{eq:opbbb}
   {\cal B}_{S\perp}^{n\mu}  &= 
      \frac{1}{g}\, \big[ S_n^\dagger i D_{S\perp}^\mu S_n \big] =
      \frac{1}{g}\, \frac{1}{ n\cdot \cP} \, S_n^\dagger \big[ i  n\cdot D_S\,, i D_{S\perp}^\mu \big] S_n 
      \,, \nn\\
  {\cal B}_{S\perp}^{{\bar n}\mu}  &={\bar n}
      \frac{1}{g}\, \big[ S_{\bar n}^\dagger i D_{S\perp}^\mu S_{\bar n} \big] =
      \frac{1}{g}\, \frac{1}{{\bar n} \cdot \cP} \, S_{\bar n}^\dagger \big[ i{\bar n}\cdot D_S \,, i D_{S\perp}^\mu \big] S_{\bar n} 
      \,, 
\end{align}
and adjoint
\begin{align}  \label{eq:Bmatrix}
  \widetilde {\cal B}_{n\perp}^{AB} &= - i f^{ABC}  {\cal B}_{n\perp}^C 
  \,,
  & \widetilde {\cal B}_{S\perp}^{nAB} &= - i f^{ABC}  {\cal B}_{S\perp}^{nC} .
\end{align}

The Glauber action includes operators that mediate soft collinear scattering 

\begin{align}  \label{eq:Ons}
    O^{qq}_{ns} &=  {\cal O}_n^{q B} \frac{1}{\cP_\perp^2}  {\cal O}_s^{q_n B} \,,
   & O^{qg}_{ns} &= {\cal O}_n^{q B} \frac{1}{\cP_\perp^2}  {\cal O}_s^{g_n B} \,,
   & O^{gq}_{ns} &=  {\cal O}_n^{g B} \frac{1}{\cP_\perp^2}  {\cal O}_s^{q_n B} \,,
   & O^{gg}_{ns} &= {\cal O}_n^{g B} \frac{1}{\cP_\perp^2}  {\cal O}_s^{g_n B} \,.
\end{align}

\begin{align}  \label{eq:Osqgn}
   {\cal O}_s^{q_n B} & =  8\pi\alpha_s\: \Big( \bar\psi^n_{S} \, T^{B} 
   \frac{n\!\!\!\slash}{2} \psi^n_{S} \Big)
  \,, \nn\\
   {\cal O}_s^{g_n B} & = 8\pi\alpha_s\: \Big(
  \frac{i}{2} f^{BCD}  {\cal B}_{S\perp\mu}^{n C}\, 
  \frac{n}{2} \cdot (\cP\!+\!\cP^\dagger)  {\cal B}_{S\perp}^{n D\mu} \Big)  \,.
\end{align}
$\psi_S$ are soft fermionic modes that appear in vacuum polarization graphs that dress the Glauber exchange.

The leading order Glauber action can then be written as
\begin{align}  \label{eq:LG}
  {\cal L}_G 
 & =
      \:  
        \sum_{i,j=q,g}    {\cal O}_n^{i B} \frac{1}{\cP_\perp^2} {\cal O}_s^{BC}  \frac{1}{\cP_\perp^2} {\cal O}_{\bar n}^{j C}  
   + \:  \sum_{i,j=q,g}  ( {\cal O}_n^{i B} \frac{1}{\cP_\perp^2} {\cal O}_s^{j_n B} 
   + {\cal O}_{\bar n}^{i B} \frac{1}{\cP_\perp^2} 
   {\cal O}_s^{j_{\bar n} B}) \,.
\end{align}
Note that this action is exact in $\alpha_s $ expansion \cite{Rothstein:2016bsq}.

\subsection{Factorization of the Amplitude from Glauber SCET}

To include effects of the Glaubers within the EFT, following \cite{Rothstein:2016bsq}, we start with the time evolution operator 
\begin{align}
U(a,b;T ) & = \lim_{T\to \infty (1-i0)} \int \big[{\cal D}\phi\big] \exp\bigg[ i \int_{-T}^{T}\!\! d^4x\: \big( {\cal L}_{n\bar{n} s}^{(0)}(x) + {\cal L}_G^{\rm II(0)}(x)\big) \bigg] \,,
\end{align}
one then expands in the number of Glauber potential insertions attaching to the $n$ and $\bar{n}$ projectiles, given by $i$ and $j$ respectively, so that
\begin{align}
\exp\bigg[ i \int_{-T}^{T}\!\! d^4x\: \big( {\cal L}_G^{\rm II(0)}(x)\big) \bigg]= 1 +\sum\limits_{i=1}^\infty \sum\limits_{j=1}^\infty U_{(i,j)}.
\end{align}
For any number of Glauber potential insertions, one can then factorize the soft and collinear operators to give a factorized expression for the amplitude for scattering of projectile $\kappa$ with $\kappa^\prime$ is
\bea
\label{amp}
M^{\kappa \kappa^\prime}=i\sum_{MN} \int \int_{\perp(N,M)} J_{\kappa N}^{A_1...A_N}(l_{1\perp}...l_{N\perp},\epsilon,\eta) 
S^{A_1..A_N;B_1..B_M}_{ (N,M)}(l_{1\perp}...l_{N\perp};l^\prime_{1\perp}...l^\prime_{M\perp})
\bar J_{\kappa^\prime M}^{B_1...B_N}(l^\prime_{1\perp}...l^\prime_{M\perp},\epsilon,\eta) \nn \\
\eea
where, following the notation in \cite{Gao2024}, we defined
\bea
\label{conv}
\int \int_{\perp(N,M)}= \frac{(-i)^{N+M}}{N! M!} \int \prod_{i=1}^N\prod_{j=1}^M
 \frac{d^{d^\prime}l_{i\perp}}{l^2_{i\perp}}\frac{d^{d^\prime}l^\prime_{j\perp}}{l^{\prime 2}_{j\perp}}\delta^{d^\prime}(\sum l_{i\perp}-q_\perp )\delta^{d^\prime}(\sum {l^\prime}_{j\perp}-q_\perp).
\eea
$\kappa$ and $\kappa^\prime$ label the external states, i.e. quarks or gluons.
Note that in Eq.(\ref{amp}) all of the Glauber light cone momentum integrals have been performed.
 since these momenta components scale as $\lambda^2$ they are dropped
in the soft loops according to the multipole expansion necessary to maintain manifest power counting.
Thus all of the Glauber loops correspond to box integrals which are rapidity finite and give a result
independent of the perp momenta. After performing the Glauber energy integral by contours we then use
\beq
\label{gint}
\int \frac{[dk_z]}{-2k_z+A +i \epsilon}\mid \frac{2 k_z}{\nu} \mid^{-\eta}=-\frac{1}{4}.
\eeq
More generally it was shown in \cite{Rothstein:2016bsq} that  the n-Glauber box diagram generates a factor of $\frac{i^n}{n!}$ as
the sum of the boxes generates the semi-classical phase. This explains why the amplitude is defined with 
the factorial prefactors in Eq.(\ref{conv}).

The jet function are defined as time order products, e.g. at the one and two Glauber gluon level
\bea
J^{A_1}(k_\perp)&=&\int dx^\pm_1  \langle p \mid T( (O^{qA_1}_n+O^{gA_1}_n)(k_\perp,x_1^\pm) \mid p^\prime \rangle \nn \\
J^{A_1 A_2}(k_\perp,k^\prime_\perp)&=&\int dx^\pm_1 dx^\pm_2 \langle p \mid T( (O^{qA_1}_n+O^{gA_1}_n)(k_\perp,x_2^\pm)  (O^{qA_2}_n+O^{gA_2}_n)(k_\perp,x_1^\pm)  \mid p^\prime \rangle,
\eea
The jets are written in this way because the combination $(O^{qA_1}_n+O^{gA_1}_n)$ is an eigenvector of $\nu\frac{d}{d\nu}$ as
shown in \cite{Rothstein:2016bsq}.

At tree level we have
\bea
\label{treeJ}
J_q^{(0)A_1...A_N}&=&g^N \bar u_n T^{A_1}...T^{A_N} \frac{n \!\!\! \slash}{2} u_n \nn \\
J_g^{(0)A_1...A_N}&=& g^N \epsilon^{\star \mu} \epsilon^{\star \nu} b_{\mu \nu} {\cal T}^{A_1}... {\cal T}^{A_N}.
\eea
where $b^{\mu \nu}= g^{\mu \nu}_\perp \bar n \cdot p_1-\bar n^\mu p_1^\nu -\bar n^\mu p_4^\nu +\frac{p_1^\perp \cdot p_4^\perp}{\bar n \cdot p_4}\bar n^{\mu}\bar n^\nu$ and $\bar n \cdot p_1 = \bar n \cdot p_4$.
Note that the intermediate propagtor is gone because it is eikonal and is replaced by unity once the light-cone momentum integral has been performed using Eq.(\ref{gint}).

While the tree level soft function is given by
\beq
\label{treeS}
S_{(i,j)}^{(0)A_1...A_i;B_1...B_j}(l_{i \perp};l_{i\perp}^\prime)=2 \delta_{ij}i^j j! \delta^{A_1 B_1}....\delta^{A_N B_N} \prod_{a=1}^j l_{i \perp}^{\prime 2}\prod_{n=1}^{j-1} \delta^{2-2\epsilon}(l_{n \perp}-l_{n \perp}^\prime).
\eeq

Each $J_i$ and $S_{ij}$ is decomposed  into irreducible representations of the $SU(N)$  symmetry and satisfy
 the RRG equation
\begin{align}
\label{RRG}
    \nu\frac{\partial}{\partial\nu}J_{\kappa(i)} &= \sum_{j=1}^\infty J_{\kappa(j)}\otimes\gamma^J_{(j,i)},\nonumber\\
    \nu\frac{\partial}{\partial\nu}S_{(i,j)}& =- \sum_{k = 1}^\infty \gamma^S_{(i,k)}\otimes S_{(k,j)} - \sum_{k = 1}^\infty S_{(i,k)}\gamma^S_{(k,j)}, \\
    \nu\frac{\partial}{\partial\nu}\bar{J}_{\kappa'(i)} &= \sum_{j=1}^\infty \gamma^{\bar J}_{(i,j)}\otimes \bar{J}_{\kappa'(j)}.\nonumber
\end{align}

In general, operators with different numbers of Glaubers, but in the same color irrep, will mix (for a discussion of the general structure see \cite{Gao2024}).
 The  single Glauber exchange is pure $8_A$ (so there is no need to label the color irrep for $\gamma_{(1,1)}$) and leads to
 a pure Regge pole, i.e. $\gamma_{(1,1)}$ is multiplicative (as opposed to convolutive), while 
 Multi-Glauber exchange also generates an $8_A$
 contribution, which in general leads to a Regge cut (i.e. a convolutive renormalizaiton group equation). 

The  full RRG system is truncated based on the fact that each Glauber is accompanied by a power of $\alpha$.
Note that, in general, $i\neq j$ in the above factorization, i.e.  there are transitions from $i$ to $j$ Glauber exchanges through the soft function. 
While the rapidity running of this system is quite nettlesome, there are significant simplifications. 
Not surprisingly, by symmetry $\gamma_J= \gamma_{\bar J}$.  Furthermore, the anomalous dimensions for
the collinear and soft pieces are not independent as there is a consistency condition \cite{Chiu2012ATheory} which follows from the
fact that the full amplitude is free of rapidity divergences, which are an artifact of factorization.  In the present case, due to
the mixing structure, this condition can be written in terms of a matrix equality \cite{Gao2024} $(Z_S)_{ij}= (Z^{-1}_J)_{ij}$.

\section{Implications of Eq.(\ref{Fadin})}

The form of the amplitude (\ref{Fadin}) implies that  the form rapidity RG equations (\ref{RRG}) in the EFT
must be strongly constrained at least up to NLL and to all orders in the planar limit. 
At NLL, where we keep all terms which scale as $\alpha^2 \log(s/t)$, with $\alpha \log(s/t)\sim 1$,  we need to run the one Glauber operator at two loops and the octet of the  two Glauber operator at one.

The form of  (\ref{Fadin}) implies that the RRG equation (in the $8_A$ channel) has  a power law solution $(s/t)^\alpha$.  This is prima facia consistent with the one Glauber exchange where the factorized form will reduce to a product (Regge Pole) since the exchanged momentum is fixed to be $q_\perp=\sqrt{-t}$. However, once we allow for two Glauber exchange we expect a convolution (Regge cut). Moreover, in principle, we can have mixing between one and two Glauber exchanges, i.e. $\gamma^{8_A}_{1,2}\neq 0$\footnote{There is no reason in general why non-planar diagrams could not generate cut structures 
either in the next to leading order two Glauber anomalous dimension or the leading order three Glauber anomalous dimension.}.
Finally, even if the two Glauber exchange RRG had a power law solution, the form of $(\ref{Fadin})$ implies that integral of  $\gamma_{(2,2)}$ has  to be fixed  by $\gamma_{(1,1)}$.
This also implies that there can be no mixing between the one and two Glauber sectors of the theory, e.g. $\gamma_{(1,2)}=0$
Calculating in the EFT it has been found that \cite{Gao2024}, not only is $\gamma_{(1,2)}=0$, but {\it all transitions }between one and multi-Glauber exchanges vanishes. i.e. beyond the one and two Glauber sector.

If we go to the planar limit this leads to an infinite number of constraints since the solution to the RG  have a pole structure, i.e no convolutions, such that the RRG has a simple exponential solution as in Eq. (\ref{Fadin}). Moreover, there is no mixing between the different Glauber 
sectors to all order in perturbation theory.  From here on out we would like to keep things as general as possible and will
not assume planarity.


We gain further control of the RRG structure in the planar limit where Eq.(\ref{Fadin}) is valid to all orders, which implies
that the full RRG system is multiplicative, i.e. its a pure pole. Moreover all of the anomalous dimensions $\gamma_{n,n}$ 
are fixed in terms of $\gamma_{(1,1)}$. It has been known for a long time that the planar limit leads to
pure Regge poles \cite{Mandelstam:1963cw}, and how this phenomena occurs in the EFT was explained in \cite{Gao2024}.
Here we briefly summarize the arguments.  In the EFT the basic reason for the all-order pole structure is remarkably simple and it has to do with the
so-called ``collapse rule'' put forth in \cite{Rothstein:2016bsq} which states that a Glauber burst can not be interrupted. 
In figure (1) a  collinear  interruption between Glauber bursts (meaning multiple Glauber exchange with possible soft interactions) leading to a vanishing contribution. Figure (3) is planar and non-vanishing and  leads to pole behavior, while figure (2) is
non-planar and leads to cut behavior.
Thus we conclude that, in accordance  with Eq.(\ref{Fadin}), the collapse rule implies there can be no convolution, when considering the RRG for $J$ or $\bar J$, 
since all the Glauber loop momenta can be run through a single collinear line.  

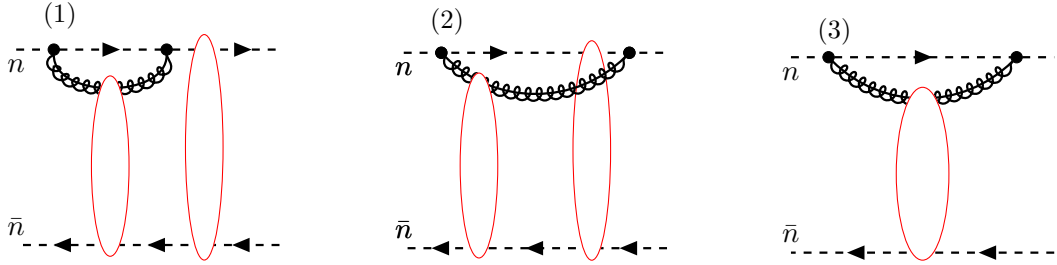
\begin{figure}
\renewcommand\thesubfigure{\arabic{subfigure}}
\begin{subfigure}[b]{0.31\textwidth}
\caption{\qquad \qquad \qquad \qquad }
\centering
\scalebox{1}{
\begin{tikzpicture}
\begin{feynman}
	\vertex [label = below: \(n\)] (i1)  at (-1.75,1.1);
	\vertex  [label = below: \(\)] (f1) at (1.75, 1.1);
	\vertex [label = above: \(\bar{n}\)] (i2)  at (-1.75,-1.5);
	\vertex [label = above: \(\)]  (f2) at (1.75, -1.5);
	\node[dot] (cn1) at (-1.25,1.1);
	\node[dot] (cn2) at (.25,1.1);
	\vertex (g11) at (-.5,.6);
	\vertex (g12) at (.75,1.1);
	\vertex (g21) at (-.5,-1.5);
	\vertex (g22) at (.75,-1.5);
	\vertex (c1) at (0, 1.75);
	\vertex (c2) at (0,-1.75);
    \vertex (c3) at (0, 1.1);
	\vertex (c4) at (0, -1.5);
\diagram*{
	(i1)--[ scalar, line width = 0.3mm](cn1)--[charged scalar, line width = 0.3mm](cn2)--[scalar, line width = 0.3mm](g12)--[charged scalar, line width = 0.3mm](f1),
	(f2)--[charged scalar, line width = 0.3mm](g22)--[charged scalar, line width = 0.3mm](g21)--[charged scalar, line width = 0.3mm](i2),
	(g11)--[scalar, red, line width = 0.3mm](g21),
	(g12)--[scalar, red, line width = 0.3mm](g22),
(cn1)--[line width = 0.3mm, half right, looseness = 1](cn2),
(cn1)--[gluon, line width = 0.3mm, half right, looseness = 1](cn2),
};
\end{feynman}
	\filldraw[red] (-.5,.6) ellipse (0.6mm and 1.2mm);
	\filldraw[red] (-.5,-1.5) ellipse (0.6mm and 1.2mm);
	\filldraw[red] (.75,1.1) ellipse (0.6mm and 1.2mm);
	\filldraw[red] (.75, -1.5) ellipse (0.6mm and 1.2mm);
	\filldraw[white](-.5,-.45) ellipse (2.5mm and 12mm);
	\draw[red](-.5,-.45) ellipse (2.5mm and 12mm);
	\filldraw[white](.75,-.2) ellipse (2.5mm and 15mm);
	\draw[red](.75,-.2) ellipse (2.5mm and 15mm);
\end{tikzpicture}
}
\end{subfigure}
\begin{subfigure}[b]{0.31\textwidth}
\caption{\qquad\qquad\qquad\qquad}
\centering
\scalebox{1}{
\begin{tikzpicture}
\begin{feynman}
	\vertex [label = below: \(n\)] (i1)  at (-1.75,1.1);
	\vertex  [label = below: \(\)] (f1) at (1.75, 1.1);
	\vertex [label = above: \(\bar{n}\)] (i2)  at (-1.75,-1.5);
	\vertex [label = above: \(\)]  (f2) at (1.75, -1.5);
	\node[dot] (cn1) at (-1.25,1.1);
	\node[dot] (cn2) at (1.25,1.1);
	\vertex (m1) at (0, 1.1);
	\vertex (m2) at (0,-1.5);
	\vertex (g11) at (-.75,.7);
	\vertex (g12) at (.75,1.1);
	\vertex (g21) at (-.75,-1.5);
	\vertex (g22) at (.75,-1.5);
	\vertex (c1) at (0, 1.75);
	\vertex (c2) at (0,-1.75);
    \vertex (c3) at (0, 1.1);
	\vertex (c4) at (0, -1.5);
\diagram*{
	(i1)--[charged scalar, line width = 0.3mm](g12)--[scalar, line width = 0.3mm](f1),
	(f2)--[charged scalar, line width = 0.3mm](g22)--[charged scalar, line width = 0.3mm](g21)--[charged scalar, line width = 0.3mm](i2),
	(g11)--[scalar, red, line width = 0.3mm](g21),
	(g12)--[scalar, red, line width = 0.3mm](g22),
};
\end{feynman}
	\filldraw[red] (-.75,.7) ellipse (0.6mm and 1.2mm);
	\filldraw[red] (-.75,-1.1) ellipse (0.6mm and 1.2mm);
	\filldraw[red] (.75,1.1) ellipse (0.6mm and 1.2mm);
	\filldraw[red] (.75, -1.1) ellipse (0.6mm and 1.2mm);
	\filldraw[white](.75,  -0.2) ellipse (2.5mm and 14.6mm);
	\draw[red](.75,  -0.2) ellipse (2.5mm and 14.6mm);
\begin{feynman}
	\vertex [label = below: \(n\)] (i1)  at (-1.75,1.1);
	\vertex  [label = below: \(\)] (f1) at (1.75, 1.1);
	\vertex [label = above: \(\bar{n}\)] (i2)  at (-1.75,-1.5);
	\vertex [label = above: \(\)]  (f2) at (1.75, -1.5);
	\node[dot] (cn1) at (-1.25,1.1);
	\node[dot] (cn2) at (1.25,1.1);
	\vertex (m1) at (0, 1.1);
	\vertex (m2) at (0,-1.5);
	\vertex (g11) at (-.75,.7);
	\vertex (g12) at (.75,1.1);
	\vertex (g21) at (-.75,-1.5);
	\vertex (g22) at (.75,-1.5);
	\vertex (c1) at (0, 1.75);
	\vertex (c2) at (0,-1.75);
    \vertex (c3) at (0, 1.1);
	\vertex (c4) at (0, -1.5);
\diagram*{
(cn1)--[line width = 0.3mm, quarter right, looseness = 1](cn2),
(cn1)--[gluon, line width = 0.3mm, quarter right, looseness = 1](cn2),
};
\end{feynman}
\filldraw[white](-.75,  -0.4) ellipse (2.5mm and 12.3mm);
	\draw[red](-.75,  -0.4) ellipse (2.5mm and 12.3mm);
\end{tikzpicture}
}
\end{subfigure}
\begin{subfigure}[b]{0.31\textwidth}
\caption{\qquad\qquad\qquad\qquad}
\centering
\scalebox{1}{
\begin{tikzpicture}
\begin{feynman}
	\vertex [label = below: \(n\)] (i1)  at (-1.75,1.1);
	\vertex  [label = below: \(\)] (f1) at (1.75, 1.1);
	\vertex [label = above: \(\bar{n}\)] (i2)  at (-1.75,-1.5);
	\vertex [label = above: \(\)]  (f2) at (1.75, -1.5);
	\node[dot] (cn1) at (-1.25,1.1);
	\node[dot] (cn2) at (1.25,1.1);
	\vertex (g11) at (-.75,.75);
	\vertex (g12) at (.75,.75);
	\vertex (g21) at (-.75,-1.5);
	\vertex (g22) at (.75,-1.5);
	\vertex (c1) at (0, 1.75);
	\vertex (c2) at (0,-1.75);
    \vertex (c3) at (0, 1.1);
	\vertex (c4) at (0, -1.5);
\diagram*{
	(i1)--[scalar, line width = 0.3mm](cn1)--[charged scalar, line width = 0.3mm](cn2)--[scalar, line width = 0.3mm](f1),
	(f2)--[charged scalar, line width = 0.3mm](c4)--[charged scalar, line width = 0.3mm](i2),
(cn1)--[line width = 0.3mm, quarter right, looseness = 1](cn2),
(cn1)--[gluon, line width = 0.3mm, quarter right, looseness = 1](cn2),
};
\end{feynman}
	\filldraw[white](0,-.45) ellipse (3.5mm and 11.5mm);
	\draw[red](0,-.45) ellipse (3.5mm and 11.5mm);
\end{tikzpicture}
}
\end{subfigure}
\caption{Different configurations of Glauber Bursts.  Here the bubble outlined in red represents an arbitrary number of Glauber exchanges.  The first diagram presents a planar configuration of two Glauber bursts, which are interrupted by the collinear gluon-quark vertex.  This class of diagrams all vanish by the collapse rule.  Diagram (2.) represents a non-planar set of Glauber exchanges which are in general non-vanishing and will generate non-pole solutions.  Lastly Diagram (3.) is a planar Glauber burst which is not interrupted.  }
\end{figure}

\section{Explaining the Iterative Structure}

As mentioned in the introduction an interesting piece of data regarding the perturbative series is that the
finite part of the two loop anomalous dimensions can be extracted from the $O(\epsilon)$ piece of the
one loop calculations.
By comparing this form to  Eq.(\ref{Fadin}) the authors of  \cite{Moult:2022lfy} were able to derive a relation between the two loop
Regge trajectory and  the constant pieces of the one and two Glauber exchange graphs:
 $S_{2,2}^{(1)}$, 
$S_{2,2}^{(0)}$, $S_{1,1}^{(1)}$,  in the color octet channel, where the superscript denotes
the order of the correction beyond the Glauber exchange. 
In particular, they derived the relation
\bea
\alpha^{(1)} &=& -2 S_{2}^{(0)}\\
\alpha^{(2)} &=&-2\l(S_{1}^{(1)} 
- S^{(1)}_{1}
S^{(0)}_{2}
\r),
\eea
where  the $S_i^{(n)}$ are defined via
\bea
S_{(1,1)}^{8_A}&=&S_1^{(0)}(1+\tilde \alpha S_1^{(1)}+\tilde \alpha^2 S_1^{(2)}+.....) \nn \\
S_{(2,2)}^{8_A}&=&i\pi S_1^{(0)}(\tilde \alpha S_2^{(0)}+\tilde \alpha^2 S_2^{(1)}+.....)\nn \\
 S_1^{(0)}&=&\frac{8 \pi i\alpha_s}{t}
 \eea
 and
\beq
\alpha(t)= \sum_i  \alpha^{(i)} \l(\frac{\tilde \alpha}{4\pi}\r)^i,
\eeq
where $\tilde \alpha$ is the rescaled coupling constant (see \cite{Moult2022AnomalousConstants}).
Not only is this useful to explain why the 
two loop Regge trajectory shows up in a one loop calculation, but it also simplifies the calculation
of the two loop trajectory itself,  as it drastically reduces the number of diagrams one needed to calculate,
since we don't need to consider two loop soft corrections to one Glauber exchange ($S_{1,1}^{(2)}$), and  only need to calculate three diagrams \cite{Moult:2022lfy}.
Moreover, these three two loop diagrams all involve a Glauber loop  which reduces the relativistic loop order by one.

In this paper we will utilize the recently developed formalism \cite{rothstein2024extracting} to  generalize the  results in \cite{Moult:2022lfy} to all orders and in any color channel. Moreover, the formalism will also generate relations between anomalous dimensions of operators
with differing number of Glaubers.

\section{Unitarity Methods }

\subsection{The general methodology}.

The main tool we will utilize in generating results will be the formalism introduced in  \cite{rothstein2024extracting} which demonstrated how one can use unitarity methods  \cite{Caron-Huot:2016cwu} to calculate rapidity anomalous dimensions efficiently.   In   \cite{rothstein2024extracting}  the rapidity anomalous dimension is calculated using cut diagrams which effectively
reduces the loop order by one and  eliminates the need for a cumbersome rapidity regulator.
Here we will briefly review how this method is used to calculate rapidity anomalous dimensions for form factors before going on
to the case of  Regge kinematics scattering amplitudes.

The basic idea introduced in \cite{Caron-Huot:2016cwu} is to extract logarithms from calculating the imaginary part since
the imaginary part is generated from the discontinuity across the branch cut. Thus one can track the log from the calculation
of the imaginary piece.  This is accomplished by acting on the matrix element with a generator in such a way as to
 rotate the argument of the log by a $2\pi$ phase, which effectively causes a crossing of the branch cut.
In the case of RG logs this rotation is accomplished by a dilatation such that one generates the fundamental relation
\beq
e^{-i \pi D} {\cal F}^\star={\cal S} {\cal F}^\star
\eeq
where $D$ is the generator of dilatations, $S=1+iM$ and ${\cal F}$ is the matrix element of the relevant form factor whose
anomalous dimension one intends to extract.
The dilation acts on the momenta $p\rightarrow e^{-i \pi}P$ such that the Mandelstam invariants are rotated in the complex plane by $2\pi$.
 Given that the  operator is an eigenvector of $D$ with eigenvalue $\gamma_\mu$
one can generate a relation between the anomalous dimensions $\gamma_\mu$ and cut graphs. For instance, at one loop we have the relation
\beq
-\pi \gamma_\mu {\cal F}^\star= {\cal M}^{(0)} {\cal F}^{(1)\star}+ {\cal M}^{(1)} {\cal F}^{(0)\star}.
\eeq 
The numerical superscript denote the order in perturbation theory.
The RHS of this equation is calculated via cut graphs.
\subsection{The Extension to Rapidity Logs}
\label{extensions}
This method won't work directly for the case of rapidity logs because their arguments are ratios of kinetmatic variables, i.e.
they dont involve the UV regulator so that a dilatation will leave the argument of the log unchanged. As such, acting with dilatations will not  be  efficacious. 
Instead  following \cite{rothstein2024extracting} we act with a modified
complex boost
  defined via
\beq
\bar K_z \equiv \sum_{i = n, \bar{n}}K_z^i,
\eeq
where $K_z^i$ is the boost in the $i$th collinear sector's $z$-direction,
\bea
K_z^n = \sum_{\{p_j\in n\}}\left(p_j^+\frac{\partial}{\partial p_j^+} - p_j^-\frac{\partial}{\partial p_j^-}\right),\\
K_z^{\bar{n}} =  \sum_{\{p_j\in \bar{n}\}}\left(p_j^-\frac{\partial}{\partial p_j^-} - p_j^+\frac{\partial}{\partial p_j^+}\right).\nn
\eea
 whose action on the light-cone coordinate momenta is
\bea
p_n^\mu &= (p_n^+,\, p_n^-,\, p_\perp^\mu) \rightarrow (e^{ \gamma}p_n^+,\, e^{- \gamma}p_n^-,\, p_{n\perp}^\mu) \\
p_{\bar{n}}^\mu  &= (p_{\bar{n}}^+,\, p_{\bar{n}}^-,\, p_{\bar{n}\perp}^\mu) \rightarrow (e^{- \gamma}p_{\bar{n}}^+,\, e^{ \gamma}p_{\bar{n}}^-,\, p_{\bar{n}\perp}^\mu),\nn
\eea
which corresponds to distinct boosts in the $n$ and $\bar n$ directions.
The action on the momentum invariants is then
\begin{equation}
    s = p_n^+ p_{\bar{n}}^-\rightarrow e^{2\gamma} s, \qquad  u = - p_n^+ p_{\bar{n}}^-\rightarrow e^{2\gamma} u,\qquad t\rightarrow t.
\end{equation}

To extract the rapidity anomalous dimensions one can work in EFT where one introduces a rapidity factorization scale $\nu$.
While the full amplitude is independent of $\nu$, the factorized pieces will depend upon $\nu$.

For two to two scattering the factorized amplitude takes the generic  form
\beq
iM= J \otimes S \otimes \bar J
\eeq
where $\otimes$ stands for a convolution in transverse momenta, $J$ and $\bar J$ are functions of collinear modes and $S$ is a soft function.
Rapidity anomalous dimensions  ($\gamma_\nu$) are
defined  via
\beq
\nu \frac{d}{d\nu}(S,J)=(\gamma_S S,\gamma_J J).
\eeq
The $\nu$  independence of the amplitude imposes the constraint mentioned in the introduction
\beq
\gamma_S +\gamma_J+\gamma_{\bar J}=0.
\eeq
Then following arguments similar to those for the canonical RG anomalous dimension discussed above one generates the master formula
\beq
e^{-i\pi(\gamma_\nu^J+\gamma_\nu^{\bar J})}{\cal F}^\star=(1+{\cal M}){\cal F}^\star.
\eeq
Note that the soft anomalous dimensions does not appear in this equation because the soft function is
independent of the hard collinear momentum upon which the boost acts.  In \cite{rothstein2024extracting} this result was used to
calculate the two loop anomalous dimensions for the Sudakov form factor as well as for the transverse momentum
soft function.
\subsection{Application to Regge Kinematics}

We would now like to apply this methodology to the present case of interest, near forward scattering.
The use of unitarity in this regime is certainly not new, and goes back to, at least,  \cite{Kuraev:1976ge}.
However, here we will  exploit unitary in a systematic fashion within the confines of an EFT to generate relations
for anomalous dimensions to all order in pertrubation theory.
If we try to apply the same arguments as we did above, regarding correlations between the logs and the phase, to this case, it becomes clear that a straight forward application will fail. To understand this we recall that the near forward scattering process is semi-classical and
as such the amplitude comes with a phase $e^{iS_{cl}}$ which is not  directly associated with the quantum rapidity logs. 
This can be most straightforwardly seen in QED, where the photon does not Reggeize, but there is certainly a phase which
arises from the semi-classical solution.  In QCD there is a quantum rapidity log  that Reggeizes the gluon, but its' relationship to the $\i \pi$ that arises do to the Glauber loop, at face value, is unclear. Consider  quark-antiquark scattering, with a gluon mass regulator $m$, we have
\bea
\label{1loop}
\text{1-loop} 
&=& \frac{ i \alpha_s^2}{t} S_{n\bar{n}}^{(1)} \left[ 8 i \pi \ln \left(  \frac{-t}{m^2} \right) \right] +  \frac{i \alpha_s^2}{t} S_{n\bar{n}}^{(2)} \left[ - 4 \ln^2 \left( \frac{m^2}{-t} \right) - 12 \ln \left( \frac{m^2}{-t} \right) - 14 \right] \nonumber \\ &+&  \frac{i \alpha_s^2}{t} S_{n\bar{n}}^{(3)} \left[ - 4 \ln \left( \frac{s}{-t} \right) \ln \left( \frac{-t}{m^2} \right) + \frac{22}{3} \ln \left( \frac{\mu^2}{-t} \right) + \frac{170}{9} + \frac{2 \pi^2}{3} \right]\nonumber \\ &+&  \frac{i \alpha_s^2}{t} S_{n\bar{n}}^{(4)} \left[ - \frac{8}{3} \ln \left( \frac{\mu^2}{-t} \right) - \frac{40}{9} \right], \nonumber \\
\eea
where the various color and spinor prefactors are given by
\bea
S_{n\bar{n}}^{(1)} &=& -\left[\bar{u}_n T^A T^B \frac{\not{n}}{2} u_n \right]\left[\bar{v}_{\bar{n}} \overline{T}^A \overline{T}^B \frac{\not{\bar{n}}}{2} v_{\bar{n}}\right],~~~~
S_{n\bar{n}}^{(2)} = C_F \left[\bar{u}_n T^A \frac{\not{\bar{n}}}{2} u_n \right]\left[\bar{v}_{\bar{n}}\bar T^A \frac{\not{n}}{2} v_{\bar{n}}\right],
\nonumber \\
S_{n\bar{n}}^{(3)} &=& C_A \left[\bar{u}_n T^A \frac{\not{\bar{n}}}{2} u_n \right]\left[\bar{v}_{\bar{n}} \bar T^A \frac{\not{n}}{2} v_{\bar{n}}\right],
~~~~
S_{n\bar{n}}^{(4)} = T_F n_f \left[\bar{u}_n T^A \frac{\not{\bar{n}}}{2} u_n \right]\left[\bar{v}_{\bar{n}} \bar T^A \frac{\not{n}}{2} v_{\bar{n}}\right].
\eea
 We see that the $i\pi$ does not come with the same color structure as the rapidity $\log(s)$, which is inherently non-Abelian and thus we must consider an object which directly ties the phase to the rapidity logs.

\subsection{Amplitudes of definite signature}
We can get hint as to the proper direction by first recalling that the $i\pi$ in Eq. (\ref{1loop}) arises from the box Glauber diagram. Moreover, the
cross-boxed diagram, which would carry a non-Abelian color factor and could correlate with the rapidity log,  vanishes in the effective theory (see section 5 of \cite{Rothstein:2016bsq}).
However, if we consider the crossed-amplitude ($s\rightarrow u=-s+O(t/s)$), the box will come with a non-Abelian color factor. In fact,
if we consider the linear combination $M_s-M_u$ the Glauber contribution will generate exactly the $C_A$ color factor
that must appear with the rapidity log. This combination is what is known as a ``negative signature amplitude" and it (along with the positive signature case)
have been an object of study in the forward scattering amplitude for many decades (see e.g. \cite{DelDuca:2018nsu}).

Much can be gleaned about the definite signature amplitudes from dispersion relations. In particular, it has been pointed out
 \cite{Caron-Huot2017Two-partonLimit} that  these objects have exactly the reality properties needed to fully control the phase from the logarithms.  Generally, this is used to simplify calculations by allowing one to drop/ignore $i\pi$ terms that might complicate the calculation.  Here we are able to exploit this connection by combining signature with the factorization of the amplitude into multi-Glauber operators.
 One can decompose the 2-to-2 amplitude as 
\begin{align}
\mathcal{M}_{2\rightarrow 2} &= \mathcal{M}^{(-)} + \mathcal{M}^{(+)},\nonumber\\
\mathcal{M}^{(\pm)}&= \frac12\left(\mathcal{M}_{2\rightarrow 2}(s,t) \pm \mathcal{M}_{2\rightarrow 2}(u,t) \right),
\end{align}
where $\mathcal{M}_{2\rightarrow 2}(s,t)$ is the $s$-channel amplitude and $\mathcal{M}_{2\rightarrow 2}(u,t)$ is the $u$-channel amplitude.  Therefore the  $(+)$ and  $(-)$ signature amplitudes are even or odd under $(s\leftrightarrow u)$ crossing.  Both are functions of the crossing symmetric combination of logarithms\cite{Caron-Huot2017Two-partonLimit},
\begin{equation}
    L = \frac12\left(\log\frac{-s - i\epsilon}{-t} + \log\frac{-u - i\epsilon}{-t} \right) = \log\frac{\abs{s}}{\abs{t}} - \frac{i\pi}{2},
\end{equation}
where we have used $u = -s $ in the high-energy limit.  It was further shown in \cite{Caron-Huot2017Two-partonLimit} that the coefficients of $L$ are either purely real for the odd-signature amplitude or purely imaginary for the even-signature amplitude.  These reality properties then allow us to exactly relate the logarithms to the phase.


\subsection{Signature Symmetry and the Complex Boost}

In order to make use of the connection between the rapidity logs and the $i\pi$'s, we now utilize  the boost operator $\bar{K}_z$, defined in section (\ref{extensions}).  Then, by acting on $L$ with  $e^{ i\pi \bar K_z}$, we have
\begin{align}
    e^{i\pi \bar{K}_z}L =\frac12\left(\log\frac{-s e^{2\pi i} - i\epsilon}{-t} + \log\frac{-u - i\epsilon}{-t} \right) = \log\frac{\abs{s}}{\abs{t}} + \frac{i\pi}{2} = L^\star.
\end{align}
The boost only acts non-trivially on the $\log(-s)$ the log is evaluated right below the branch cut, while the $\log(-u)$ is evaluated away from the branch cut and a rotation by $2\pi i $ has no effect.  The definite signature amplitudes then transform as 
\begin{equation}
    e^{i\pi \bar{K}_z}\mathcal{M}^{(\pm)}(L) = \mathcal{M}^{(\pm)}(L^\star) = \mp \mathcal{M}^{(\pm)}(L)^\star.
    \label{Signature Conjugation}
\end{equation}
The last equality follows from the reality properties of the amplitudes and the Schwartz reflection principle.  

\subsection{The Master Formulae}

We now use the factorization to write the definite-signature amplitudes as
\begin{align}
    \mathcal{M}^{(\pm)} =i \sum_{i,j=1}^\infty\left[J^{(s)}_{\kappa(i)} \otimes S^{(s)}_{(i,j)} \otimes \bar{J}^{(s)}_{\kappa'(j)}  \pm J^{(u)}_{\kappa(i)} \otimes S^{(u)}_{(i,j)} \otimes \bar{J}^{(u)}_{\kappa'(j)} 
    \right],
\end{align}
where the $(s/u)$ superscripts denote whether the matrix element is computed in the $s$- or $u$-channels. 
Each component, $J,S$ etc, is decomposed in terms of color irreps, whose indices we have suppressed. 
 Since the amplitude may only depend on $p_n^+$ and $p_{\bar{n}}^-$ through the rapidity logs in the collinear functions, we may write 
\begin{align}
\bar{K}_z \mathcal{M}^{(\pm)} &= i\sum_{i,j=1}^\infty\left[\left(-\frac12\nu\frac{\partial}{\partial \nu}J^{(s)}_{\kappa(i)}\right) \otimes S^{(s)}_{(i,j)} \otimes \bar{J}^{(s)}_{\kappa'(j)}  \pm \left(-\frac12\nu\frac{\partial}{\partial \nu}J^{(u)}_{\kappa(i)}\right) \otimes S^{(u)}_{(i,j)} \otimes \bar{J}^{(u)}_{\kappa'(j)} \right]\nonumber\\
& + \sum_{i,j=1}^\infty\left[J^{(s)}_{\kappa(i)} \otimes S^{(s)}_{(i,j)} \otimes \left(-\frac12\nu\frac{\partial}{\partial \nu}\bar{J}^{(s)}_{\kappa'(j)} \right) \pm J^{(u)}_{\kappa(i)} \otimes S^{(u)}_{(i,j)} \otimes \left(-\frac12\nu\frac{\partial}{\partial \nu}\bar{J}^{(u)}_{\kappa'(j)} \right)\right].
\label{Boost to RGE}
\end{align}
Using the rapidity RRG equation we have 
\begin{align}
    \mp\mathcal{M}^{(\pm)\star} &= e^{i\pi \bar{K}_z}\mathcal{M}^{(\pm)}\nonumber\\
    &= i \sum_{i,j,k,l=1}^\infty\bigg[ \left(J^{(s)}_{\kappa(i)} e^{-\otimes \frac{i}{2}\pi\gamma_{(i,j)}} \right) \otimes S^{(s)}_{(j,k)} \otimes\left(e^{-\frac{i}{2}\pi\gamma_{(k,l)}\otimes} \bar{J}^{(s)}_{\kappa'(l)} \right) \nonumber\\
    &\qquad \pm \left(J^{(u)}_{\kappa(i)} e^{-\otimes \frac{i}{2}\pi\gamma_{(i,j)}}\right)\otimes S^{(u)}_{(j,k)} \otimes\left(e^{-\frac{i}{2}\pi\gamma_{(k,l)\otimes}} \bar{J}^{(u)}_{\kappa'(l)} \right) \bigg].
    \label{conjugate to RGE}
\end{align}
Note that the action of $\bar K_z$ generates a factor of $1/2$, as compared to the action of $\bar{K}_z$ when acting on an amplitude \cite{rothstein2024extracting}.  This comes from the fact that we are looking at the signatured amplitude
and the action of the generator on the difference/sum of the two channels naturally leads
to twice the anomalous dimension.


To extract the Regge trajectory, we focus on the odd-signature amplitude. Using the vanishing of $1\rightarrow j$ transitions, we can rewrite Eq.(\ref{Signature Conjugation}) as
\begin{align}
    \mathcal{M}^{(-)}_{1} e^{-i\pi\gamma_{(1,1)}}
    = \mathcal{M}^{(-)\star}-e^{i\pi \bar{K}_z}\mathcal{M}^{(-)}_{\geq 2},
\end{align}
where we have used the fact that the one Glauber exchange is multiplicative (Regge pole).

We now use that the action of the boost is to transform $L$ to $L^\star$, or shift $\log(s)\rightarrow \log(s) + i\pi$.  With this, we can write the master formula for the Regge trajectory as
\begin{equation}
   \mathcal{M}^{(-)}_{1} e^{-i\pi\gamma_{(1,1)}} =\mathcal{M}^{(-)\star} - \left.\mathcal{M}^{(-)}_{\geq 2}\right|_{s\rightarrow  e^ {2\pi i}s}.
\end{equation}
Notice that in the second term the action of the boost generator $e^{i \pi \bar K_z}$ does not conjugate $\mathcal{M}^{(-)}_{\geq 2}$, since $L$ is formed from one Glauber (the log) and two Glauber (the $i \pi$) exchange graphs.

As a last step, we apply  the unitarity relation $({\cal M}^{(-)}- {\cal M}^{(-)\star})= i({\cal M}{\cal M}^{\star})^{(-)}$ 
where $({\cal M}{\cal M}^{\star})^{(-)}=1/2( {\cal M}_s {\cal M}_s^\star- {\cal M}_u {\cal M}_u^\star)$, we arrive at the master
formula
\begin{equation}
\boxed{
   \mathcal{M}^{(-)}_{1}\left( e^{-i\pi\gamma_{(1,1)}} -1\right) 
   = -i(\mathcal{M}\mathcal{M}^\star)^{(-)}
   +\left[\mathcal{M}^{(-)}_{\geq 2} - \left.\mathcal{M}^{(-)}_{\geq 2}\right|_{s\rightarrow  e^ {2\pi i}s}\right].
   \label{Master Regge Odd}}
\end{equation}
A few comments are in order.  First, we note that while this formula does involve more terms than just unitarity cuts (first term on the RHS), there are two major simplifications that it provides.  The first is that the terms in the square brackets always come from graphs with at least two Glauber insertions, and so must have at least one Glauber loop.  This provides many of the same technical simplifications as only working with cut graphs, as Glauber loops are easier to perform than soft or collinear loops.  The second simplification is that only terms proportional to rapidity logarithms contribute to the terms in the square brackets, as all other contributions vanish in the subtraction.  The coefficients of the rapidity logarithms are generally easier to compute than the rapidity-finite terms.
Moreover, one can also combine the two integrands (with the appropriate changes in sign in the second) to yield a rapidity finite integral that needs no regulator which can complicate the evaluation at higher loops.


Lastly, we give the equivalent formula for the even-signature sector.  Due to an extra sign in Eq. (\ref{conjugate to RGE}), we find
\begin{equation}
\boxed{
    \l(e^{i\pi \bar{K}_z}-1\r)\mathcal{M}^{(+)} = -2\text{Re}[\mathcal{M}^{(+)}].
    \label{Master Regge Even}}
\end{equation}
Unlike with the odd signature formula, the even signature relation cannot be simplified further using unitarity.
Note that $\mathcal{M}_{1}^{(+)}=0$  to all orders so to extract information from this equation we will need to go to higher orders.

\section{Calculating the Regge Trajectory through Two Loops}
\subsection{Leading order one Glauber anomalous dimension: $\gamma_{(1,1)}^{(1)}$}

We now verify the master formula in Eq. (\ref{Master Regge Odd}).  At one loop, it simplifies to
\beq
\mathcal{M}^{(-)(0)}_{1}\gamma_{(1,1)}^{(1)} = \frac{1}{\pi}\left(\mathcal{M}^\dag\mathcal{M}^{(-)}\right)^{(1)}.
\label{1 Loop Regge}
\eeq

The tree-level odd signature amplitude for quark-antiquark scattering is given by 
\begin{align}
    \mathcal{M}_1^{(-)(0)} =& \frac12\l(
\begin{gathered}
\scalebox{0.7}{
\begin{tikzpicture}
\begin{feynman}
	\vertex (f1) at (0,1);
	\vertex (f2) at (0,-1);
	\vertex[label =above: \(p_3\)] (p3) at (1.25, 1);
	\vertex[label =above: \(p_2\)] (p2) at (-1.25, 1);
	\vertex[label = below: \(p_1\)] (p1) at (-1.25, -1);
	\vertex[label = below: \(p_4\)] (p4) at (1.25, -1);
	\vertex [label= \(n\)] at (-1.25, .5);
	\vertex [label= \(\bar{n}\)] at (-1.25, -1.);
\diagram*{
	(f2)--[scalar, red, line width = 0.3mm](f1),
	(p2)--[charged scalar,  line width = 0.3mm](f1)--[charged scalar,  line width = 0.3mm](p3),
	(p4)--[charged scalar,  line width = 0.3mm](f2)--[charged scalar,  line width = 0.3mm](p1),
};
\end{feynman}
    \filldraw[red] (0,1) ellipse (0.6mm and 1.2mm);
	\filldraw[red] (0,-1) ellipse (0.6mm and 1.2mm);
\end{tikzpicture}
}
\end{gathered}
-
\begin{gathered}
\scalebox{0.7}{
\begin{tikzpicture}
\begin{feynman}
	\vertex (f1) at (0,1);
	\vertex (f2) at (0,-1);
	\vertex[label =above : \(\bar{p}_2\)] (p3) at (1.25, 1);
	\vertex[label =above : \(\bar{p}_3\)] (p2) at (-1.25, 1);
	\vertex[label =below : \(p_1\)] (p1) at (-1.25, -1);
	\vertex[label =below : \(p_4\)] (p4) at (1.25, -1);
	\vertex [label= \(n\)] at (-1.25, .5);
	\vertex [label= \(\bar{n}\)] at (-1.25, -1);
\diagram*{
	(f2)--[scalar, red, line width = 0.3mm](f1),
	(p3)--[charged scalar,  line width = 0.3mm](f1)--[charged scalar,  line width = 0.3mm](p2),
	(p4)--[charged scalar,  line width = 0.3mm](f2)--[charged scalar,  line width = 0.3mm](p1),
};
\end{feynman}
    \filldraw[red] (0,1) ellipse (0.6mm and 1.2mm);
	\filldraw[red] (0,-1) ellipse (0.6mm and 1.2mm);
\end{tikzpicture}
}
\end{gathered}
\r) 
\equiv
\left(
\begin{gathered}
\scalebox{0.7}{
\begin{tikzpicture}
\begin{feynman}
	\vertex (f1) at (0,1);
	\vertex (f2) at (0,-1);
	\vertex (p3) at (1.25, 1);
	\vertex(p2) at (-1.25, 1);
	\vertex (p1) at (-1.25, -1);
	\vertex (p4) at (1.25, -1);
	\vertex [label= \(n\)] at (-1.25, .5);
	\vertex [label= \(\bar{n}\)] at (-1.25, -1);
\diagram*{
	(f2)--[scalar, red, line width = 0.3mm](f1),
	(p2)--[charged scalar,  line width = 0.3mm](f1)--[charged scalar,  line width = 0.3mm](p3),
	(p4)--[charged scalar,  line width = 0.3mm](f2)--[charged scalar,  line width = 0.3mm](p1),
};
\end{feynman}
    \filldraw[red] (0,1) ellipse (0.6mm and 1.2mm);
	\filldraw[red] (0,-1) ellipse (0.6mm and 1.2mm);
\end{tikzpicture}
}
\end{gathered}
\right)^{(-)},\nonumber\\
&
= \left[\bar{u}_nT^A\frac{\slashed{\bar{n}}}{2}u_n\right]\frac{8\pi\alpha_s}{t}\left[ \bar{v}_{\bar{n}}\bar{T}^A\frac{\slashed{n}}{2}v_{\bar{n}}\right].
\end{align}
In the first line we have introduced the notation of taking the odd-signature piece of the diagram, which can be computed by subtracting the crossed $u$-channel diagram from the $s$-channel diagrams.  
Note that we do not need to include the contributions of the terms in the square brackets of Eq. (\ref{Master Regge Odd}) at this order, as only the tree-level matrix elements of $J_{\kappa(2)}$, $S_{(2,2)}$, and $\bar{J}_{\kappa'(2)}$ contribute, and these contain no rapidity logarithms\cite{Gao2024}.  If we compute the cut, we then find
\begin{align}
&\left(\mathcal{M}^\dag \mathcal{M}^{(-)}\right)^{(1)}=
\left(
\begin{gathered}
\scalebox{0.5}{
\begin{tikzpicture}
\begin{feynman}
	\vertex (i1)  at (-2,1.5);
	\vertex (f1) at (2, 1.5);
	\vertex (i2)  at (-2,-1.5);
	\vertex (f2) at (2, -1.5);
	\vertex (g11) at (-1,1.5);
	\vertex (g12) at (1,1.5);
	\vertex (g21) at (-1,-1.5);
	\vertex (g22) at (1,-1.5);
	\vertex (m1) at (0, 1.5);
	\vertex (m2) at (0,-1.5);
	\vertex (c1) at (0, 2.125);
	\vertex (c2) at (0,-2.125);
	\vertex [label= \(n\)] at (-2, 1);
	\vertex [label= \(\bar{n}\)] at (-2, -1.5);
\diagram*{
	(i1)--[charged scalar, line width = 0.3mm](g11)--[charged scalar, line width = 0.3mm](m1)--[charged scalar, line width = 0.3mm](g12)--[charged scalar, line width = 0.3mm](f1),
	(f2)--[charged scalar, line width = 0.3mm](g22)--[charged scalar, line width = 0.3mm](m2)--[charged scalar, line width = 0.3mm](g21)--[charged scalar, line width = 0.3mm](i2),
	(g11)--[scalar, red, line width = 0.3mm](g21),
	(g12)--[scalar, red, line width = 0.3mm](g22),
	(c1)--[scalar, gray](c2),
};
\end{feynman}
	\filldraw[red] (-1,1.5) ellipse (0.6mm and 1.2mm);
	\filldraw[red] (-1,-1.5) ellipse (0.6mm and 1.2mm);
	\filldraw[red] (1,1.5) ellipse (0.6mm and 1.2mm);
	\filldraw[red] (1, -1.5) ellipse (0.6mm and 1.2mm);
\end{tikzpicture}
}
\end{gathered}
\right)^{(-)}\nonumber\\
 &= \frac{-4\pi\alpha_s^2}{t}\left[\bar{u}_n\left(T^A T^B-T^B T^A\right)\frac{\slashed{\bar{n}}}{2}u_n\right]\left[ \bar{v}_{\bar{n}}\bar{T}^A\bar{T}^B\frac{\slashed{n}}{2}v_{\bar{n}}\right]\frac{\Gamma(1 + \varepsilon)\Gamma(-\varepsilon)^2}{\Gamma(-2\varepsilon)}\left(\frac{\bar{\mu}^2}{-t}\right)^{\varepsilon},\nonumber\\
&= \frac{C_A\alpha_s}{4}\frac{\Gamma(1 + \varepsilon)\Gamma(-\varepsilon)^2}{\Gamma(-2\varepsilon)}\left(\frac{\bar{\mu}^2}{-t}\right)^{\varepsilon} \left[\bar{u}_nT^A\frac{\slashed{\bar{n}}}{2}u_n\right]\frac{8\pi\alpha_s}{t}\left[ \bar{v}_{\bar{n}}\bar{T}^A\frac{\slashed{n}}{2}v_{\bar{n}}\right],\nonumber\\
&=\pi\gamma^{(1)}_{(1,1)}\mathcal{M}^{(-),(0)}.
\end{align}
 
Which yields the standard result for the leading order Regge trajectory
\begin{equation}
 \gamma^{(1)}_{(1,1)}= \frac{C_A\alpha_s} {4\pi} \frac{ \Gamma (-\epsilon )^2 \Gamma (\epsilon +1) \left(\frac{\bar \mu ^2}{-t}\right)^{\epsilon }}{ \Gamma (-2 \epsilon )}=-\frac{C_A\alpha_s}{2\pi}(\frac{1}{\epsilon} +\log(\frac{ \mu^2}{-t}))+O(\epsilon).
\end{equation}
Through one loop this relation is almost trivial since the amplitude is linear in $\log|s/t|$ at this order.  On the other hand, this calculation is significantly simpler than the direct calculation of the anomalous dimension (see \cite{Rothstein:2016bsq}).

\subsubsection{Next to leading order  anomalous dimension for  one Glauber operator: $\gamma_{(1,1)}^{(1)}$}

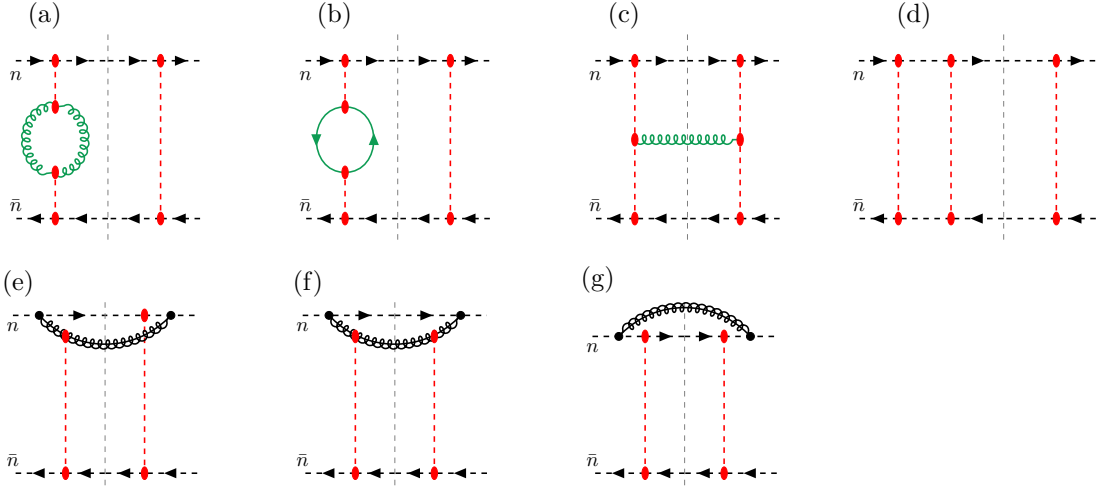
\begin{figure}
\begin{subfigure}[b]{.23\textwidth}
\caption{\qquad \qquad \qquad }
\centering
\scalebox{0.7}{
\begin{tikzpicture}
\begin{feynman}
	\vertex (f1) at (1,1.5);
	\vertex (f2) at (1,-1.5);
	\vertex (p3) at (1.75, 1.5);
	\vertex (p2) at (-1.75, 1.5);
	\vertex (p1) at (-1.75, -1.5);
	\vertex (p4) at (1.75, -1.5);
	\vertex [label= \(n\)] at (-1.75, 1);
	\vertex [label= \(\bar{n}\)] at (-1.75, -1.5);
	\vertex (g1) at (-1, 1.5);
	\vertex (g2) at (-1, -1.5);
	\vertex (c1) at (0, 2.0);
	\vertex (c2) at (0, -2.0);
 	\vertex (c3) at (0, 1.5);
	\vertex (c4) at (0, -1.5);
	\vertex (s1) at (-1,.622);
	\vertex (s2) at (-1, -.622);
\diagram*{
	(f1)--[scalar, red, line width = 0.3mm](f2),
	(p2)--[charged scalar,  line width = 0.3mm](g1)--[charged scalar,  line width = 0.3mm](c3)--[charged scalar,  line width = 0.3mm](f1)--[charged scalar,  line width = 0.3mm](p3),
	(p4)--[charged scalar,  line width = 0.3mm](f2)--[charged scalar,  line width = 0.3mm](c4)--[charged scalar,  line width = 0.3mm](g2)--[charged scalar,  line width = 0.3mm](p1),
	(c1)--[scalar, gray](c2),
	(g1)--[scalar, red, line width = 0.3mm](s1),
	(g2)--[scalar, red, line width = 0.3mm](s2),
	(s2)--[SGreen, gluon,looseness = 1.75, line width=0.3mm, half right](s1)--[SGreen, gluon,looseness = 1.75, line width=0.3mm, half right](s2),
};
\end{feynman}
	\filldraw[red] (-1,1.5) ellipse (0.6mm and 1.2mm);
	\filldraw[red] (-1,-1.5) ellipse (0.6mm and 1.2mm);
	\filldraw[red] (1,1.5) ellipse (0.6mm and 1.2mm);
	\filldraw[red] (1, -1.5) ellipse (0.6mm and 1.2mm);
	\filldraw[red] (-1,.622) ellipse (0.6mm and 1.2mm);
	\filldraw[red] (-1,-.622) ellipse (0.6mm and 1.2mm);
\end{tikzpicture}
}
\end{subfigure}
\begin{subfigure}[b]{.23\textwidth}
\caption{\qquad \qquad \qquad }
\centering
\scalebox{0.7}{
\begin{tikzpicture}
\begin{feynman}
	\vertex (f1) at (1,1.5);
	\vertex (f2) at (1,-1.5);
	\vertex (p3) at (1.75, 1.5);
	\vertex (p2) at (-1.75, 1.5);
	\vertex (p1) at (-1.75, -1.5);
	\vertex (p4) at (1.75, -1.5);
	\vertex [label= \(n\)] at (-1.75, 1);
	\vertex [label= \(\bar{n}\)] at (-1.75, -1.5);
	\vertex (g1) at (-1, 1.5);
	\vertex (g2) at (-1, -1.5);
	\vertex (c1) at (0, 2.0);
	\vertex (c2) at (0, -2.0);
	\vertex (s1) at (-1,.622);
	\vertex (s2) at (-1, -.622);
   	\vertex (c3) at (0, 1.5);
	\vertex (c4) at (0, -1.5);
\diagram*{
	(f1)--[scalar, red, line width = 0.3mm](f2),
	(p2)--[charged scalar,  line width = 0.3mm](g1)--[charged scalar,  line width = 0.3mm](c3)--[charged scalar,  line width = 0.3mm](f1)--[charged scalar,  line width = 0.3mm](p3),
	(p4)--[charged scalar,  line width = 0.3mm](f2)--[charged scalar,  line width = 0.3mm](c4)--[charged scalar,  line width = 0.3mm](g2)--[charged scalar,  line width = 0.3mm](p1),
	(c1)--[scalar, gray](c2),
	(g1)--[scalar, red, line width = 0.3mm](s1),
	(g2)--[scalar, red, line width = 0.3mm](s2),
		(s2)--[SGreen, fermion,looseness = 1.75, line width=0.3mm, half right](s1)--[SGreen, fermion,looseness = 1.75, line width=0.3mm, half right](s2),
};
\end{feynman}
	\filldraw[red] (-1,1.5) ellipse (0.6mm and 1.2mm);
	\filldraw[red] (-1,-1.5) ellipse (0.6mm and 1.2mm);
	\filldraw[red] (1,1.5) ellipse (0.6mm and 1.2mm);
	\filldraw[red] (1., -1.5) ellipse (0.6mm and 1.2mm);
	\filldraw[red] (-1,.622) ellipse (0.6mm and 1.2mm);
	\filldraw[red] (-1,-.622) ellipse (0.6mm and 1.2mm);
\end{tikzpicture}
}
\end{subfigure}
\begin{subfigure}[b]{.23\textwidth}
\caption{\qquad \qquad \qquad }
\centering
\scalebox{0.7}{
\begin{tikzpicture}
\begin{feynman}
	\vertex (f1) at (1,1.5);
	\vertex (f2) at (1,-1.5);
	\vertex (p3) at (1.75, 1.5);
	\vertex (p2) at (-1.75, 1.5);
	\vertex (p1) at (-1.75, -1.5);
	\vertex (p4) at (1.75, -1.5);
	\vertex [label= \(n\)] at (-1.75, 1.);
	\vertex [label= \(\bar{n}\)] at (-1.75, -1.5);
	\vertex (g1) at (-1, 1.5);
	\vertex (g2) at (-1, -1.5);
	\vertex (c1) at (0, 2.0);
	\vertex (c2) at (0, -2.0);
	\vertex (s1) at (1,0);
	\vertex (s2) at (-1, 0);
    \vertex (c3) at (0, 1.5);
	\vertex (c4) at (0, -1.5);
\diagram*{
	(f1)--[scalar, red, line width = 0.3mm](s1),
	(f2)--[scalar, red, line width = 0.3mm](s1),
	(p2)--[charged scalar,  line width = 0.3mm](g1)--[charged scalar,  line width = 0.3mm](c3)--[charged scalar,  line width = 0.3mm](f1)--[charged scalar,  line width = 0.3mm](p3),
	(p4)--[charged scalar,  line width = 0.3mm](f2)--[charged scalar,  line width = 0.3mm](c4)--[charged scalar,  line width = 0.3mm](g2)--[charged scalar,  line width = 0.3mm](p1),
	(c1)--[scalar, gray](c2),
	(g1)--[scalar, red, line width = 0.3mm](g2),
	(s2)--[SGreen, gluon,line width=0.3mm](s1),
};
\end{feynman}
	\filldraw[red] (-1,1.5) ellipse (0.6mm and 1.2mm);
	\filldraw[red] (-1,-1.5) ellipse (0.6mm and 1.2mm);
	\filldraw[red] (1,1.5) ellipse (0.6mm and 1.2mm);
	\filldraw[red] (1, -1.5) ellipse (0.6mm and 1.2mm);
	\filldraw[red] (1,0) ellipse (0.6mm and 1.2mm);
	\filldraw[red] (-1,0) ellipse (0.6mm and 1.2mm);
\end{tikzpicture}
}
\end{subfigure}
\begin{subfigure}[b]{.23\textwidth}
\caption{\qquad \qquad \qquad }
\centering
\scalebox{0.7}{
\begin{tikzpicture}
\begin{feynman}
	\vertex (f1) at (1.5,1.5);
	\vertex (f2) at (1.5,-1.5);
	\vertex (p3) at (2.25, 1.5);
	\vertex (p2) at (-2.25, 1.5);
	\vertex (p1) at (-2.25, -1.5);
	\vertex (p4) at (2.25, -1.5);
	\vertex [label= \(n\)] at (-2.25, 1);
	\vertex [label= \(\bar{n}\)] at (-2.25, -1.5);
	\vertex (g1) at (-1.5, 1.5);
	\vertex (g2) at (-1.5, -1.5);
	\vertex (c1) at (.5, 2.0);
	\vertex (c2) at (.5, -2.0);
 	\vertex (c3) at (-0.5, 1.5);
	\vertex (c4) at (-0.5, -1.5);
	\vertex (s1) at (.5,1.5);
	\vertex (s2) at (.5, -1.5);
\diagram*{
	(f1)--[scalar, red, line width = 0.3mm](f2),
	(p2)--[charged scalar,  line width = 0.3mm](g1)--[scalar,  line width = 0.3mm](c3)--[charged scalar,  line width = 0.3mm](s1)--[scalar,  line width = 0.3mm](f1)--[charged scalar,  line width = 0.3mm](p3),
	(p4)--[charged scalar,  line width = 0.3mm](f2)--[ scalar,  line width = 0.3mm](s2)--[charged scalar,  line width = 0.3mm](c4)--[ scalar,  line width = 0.3mm](g2)--[charged scalar,  line width = 0.3mm](p1),
	(c1)--[scalar, gray](c2),
	(g1)--[scalar, red, line width = 0.3mm](g2),
    (c3)--[scalar, red, line width = 0.3mm](c4),
};
\end{feynman}
	\filldraw[red] (-1.5,1.5) ellipse (0.6mm and 1.2mm);
	\filldraw[red] (-1.5,-1.5) ellipse (0.6mm and 1.2mm);
	\filldraw[red] (1.5,1.5) ellipse (0.6mm and 1.2mm);
	\filldraw[red] (1.5, -1.5) ellipse (0.6mm and 1.2mm);
 	\filldraw[red] (-.5,1.5) ellipse (0.6mm and 1.2mm);
	\filldraw[red] (-.5, -1.5) ellipse (0.6mm and 1.2mm);
\end{tikzpicture}
}
\end{subfigure}
\newline
\begin{subfigure}[b]{.23\textwidth}
\caption{\qquad \qquad \qquad \qquad}
\centering
\scalebox{.7}{
\begin{tikzpicture}
\begin{feynman}
	\vertex [label = below: \(n\)] (i1)  at (-1.75,1.5);
	\vertex  [label = below: \(\)] (f1) at (1.75, 1.5);
	\vertex [label = above: \(\bar{n}\)] (i2)  at (-1.75,-1.5);
	\vertex [label = above: \(\)]  (f2) at (1.75, -1.5);
	\node[dot] (cn1) at (-1.25,1.5);
	\node[dot] (cn2) at (1.25,1.5);
	\vertex (m1) at (0, 1.5);
	\vertex (m2) at (0,-1.5);
	\vertex (g11) at (-.75,1.1);
	\vertex (g12) at (.75,1.5);
	\vertex (g21) at (-.75,-1.5);
	\vertex (g22) at (.75,-1.5);
	\vertex (c1) at (0, 1.75);
	\vertex (c2) at (0,-1.75);
    \vertex (c3) at (0, 1.5);
	\vertex (c4) at (0, -1.5);
\diagram*{
	(i1)--[charged scalar, line width = 0.3mm](g12)--[scalar, line width = 0.3mm](f1),
	(f2)--[charged scalar, line width = 0.3mm](g22)--[charged scalar, line width = 0.3mm](c4)--[charged scalar, line width = 0.3mm](g21)--[charged scalar, line width = 0.3mm](i2),
	(g11)--[scalar, red, line width = 0.3mm](g21),
	(g12)--[scalar, red, line width = 0.3mm](g22),
	(c1)--[scalar, gray](c2),
(cn1)--[line width = 0.3mm, quarter right, looseness = 1](cn2),
(cn1)--[gluon, line width = 0.3mm, quarter right, looseness = 1](cn2),
};
\end{feynman}
	\filldraw[red] (-.75,1.1) ellipse (0.6mm and 1.2mm);
	\filldraw[red] (-.75,-1.5) ellipse (0.6mm and 1.2mm);
	\filldraw[red] (.75,1.5) ellipse (0.6mm and 1.2mm);
	\filldraw[red] (.75, -1.5) ellipse (0.6mm and 1.2mm);
\end{tikzpicture}
}
\end{subfigure}
\begin{subfigure}[b]{.23\textwidth}
\caption{\qquad \qquad \qquad\qquad }
\centering
\scalebox{.7}{
\begin{tikzpicture}
\begin{feynman}
	\vertex [label = below: \(n\)] (i1)  at (-1.75,1.5);
	\vertex  [label = below: \(\)] (f1) at (1.75, 1.5);
	\vertex [label = above: \(\bar{n}\)] (i2)  at (-1.75,-1.5);
	\vertex [label = above: \(\)]  (f2) at (1.75, -1.5);
	\node[dot] (cn1) at (-1.25,1.5);
	\node[dot] (cn2) at (1.25,1.5);
	\vertex (m1) at (0, 1.5);
	\vertex (m2) at (0,-1.5);
	\vertex (g11) at (-.75,1.1);
	\vertex (g12) at (.75,1.1);
    \vertex (c12) at (0,1.5);
	\vertex (g21) at (-.75,-1.5);
	\vertex (g22) at (.75,-1.5);
	\vertex (c1) at (0, 1.75);
	\vertex (c2) at (0,-1.75);
    \vertex (c3) at (0, 1.5);
	\vertex (c4) at (0, -1.5);
\diagram*{
	(i1)--[ scalar, line width = 0.3mm](cn1)--[charged scalar, line width = 0.3mm](c12)--[charged scalar, line width = 0.3mm](cn2)--[scalar, line width = 0.3mm](f1),
	(f2)--[charged scalar, line width = 0.3mm](g22)--[charged scalar, line width = 0.3mm](c4)--[charged scalar, line width = 0.3mm](g21)--[charged scalar, line width = 0.3mm](i2),
	(g11)--[scalar, red, line width = 0.3mm](g21),
	(g12)--[scalar, red, line width = 0.3mm](g22),
	(c1)--[scalar, gray](c2),
(cn1)--[line width = 0.3mm, quarter right, looseness = 1](cn2),
(cn1)--[gluon, line width = 0.3mm, quarter right, looseness = 1](cn2),
};
\end{feynman}
	\filldraw[red] (-.75,1.1) ellipse (0.6mm and 1.2mm);
	\filldraw[red] (-.75,-1.5) ellipse (0.6mm and 1.2mm);
	\filldraw[red] (.75,1.1) ellipse (0.6mm and 1.2mm);
	\filldraw[red] (.75, -1.5) ellipse (0.6mm and 1.2mm);
\end{tikzpicture}
}
\end{subfigure}
\begin{subfigure}[b]{.23\textwidth}
\caption{\qquad \qquad \qquad \qquad}
\centering
\scalebox{.7}{
\begin{tikzpicture}
\begin{feynman}
	\vertex [label = below: \(n\)] (i1)  at (-1.75,1.1);
	\vertex  [label = below: \(\)] (f1) at (1.75, 1.1);
	\vertex [label = above: \(\bar{n}\)] (i2)  at (-1.75,-1.5);
	\vertex [label = above: \(\)]  (f2) at (1.75, -1.5);
	\node[dot] (cn1) at (-1.25,1.1);
	\node[dot] (cn2) at (1.25,1.1);
	\vertex (g11) at (-.75,1.1);
	\vertex (g12) at (.75,1.1);
	\vertex (g21) at (-.75,-1.5);
	\vertex (g22) at (.75,-1.5);
	\vertex (c1) at (0, 1.75);
	\vertex (c2) at (0,-1.75);
    \vertex (c3) at (0, 1.1);
	\vertex (c4) at (0, -1.5);
\diagram*{
	(i1)--[scalar, line width = 0.3mm](g11)--[charged scalar, line width = 0.3mm](c3)--[charged scalar, line width = 0.3mm](g12)--[scalar, line width = 0.3mm](f1),
	(f2)--[charged scalar, line width = 0.3mm](g22)--[charged scalar, line width = 0.3mm](c4)--[charged scalar, line width = 0.3mm](g21)--[charged scalar, line width = 0.3mm](i2),
	(g11)--[scalar, red, line width = 0.3mm](g21),
	(g12)--[scalar, red, line width = 0.3mm](g22),
	(c1)--[scalar, gray](c2),
(cn2)--[line width = 0.3mm, quarter right, looseness = 1](cn1),
(cn2)--[gluon, line width = 0.3mm, quarter right, looseness = 1](cn1),
};
\end{feynman}
	\filldraw[red] (-.75,1.1) ellipse (0.6mm and 1.2mm);
	\filldraw[red] (-.75,-1.5) ellipse (0.6mm and 1.2mm);
	\filldraw[red] (.75,1.1) ellipse (0.6mm and 1.2mm);
	\filldraw[red] (.75, -1.5) ellipse (0.6mm and 1.2mm);
\end{tikzpicture}
}
\end{subfigure} 
\caption{Two-loop cut diagrams contributing to the two loop Regge trajectory.  Diagrams (a), (b), and (c) build up the Regge trajectory and the iterative soft contributions in the single Glauber operator, while the collinear loops (f) and (g) only build up the collinear iterative terms.  Diagrams (d) and (e) vanish when summing over their mirror images and taking the odd-signature piece.}
\end{figure}

Now we move onto the two-loop Regge trajectory, i.e. $\gamma^{(2)}_{(1,1)}$,  which is still  pure octet since we are focussing on  the one Glauber operator.  At this order  Eq. (\ref{Master Regge Odd})  becomes
\begin{align}
    \mathcal{M}^{(-)(0)}_{1}\gamma_{(1,1)}^{(2)} +\mathcal{M}^{(-)(1)}_{1}\gamma_{(1,1)}^{(1)}-&\frac{i\pi}{2!}\left(\gamma_{(1,1)}^{(1)}\right)^2\mathcal{M}^{(-)(0)}_{1}= \frac{1}{\pi}\left(\mathcal{M}^\dag\mathcal{M}^{(-)}\right)^{(2)} \\
    &\qquad\qquad\qquad   + \frac{ i}{\pi}\left[\mathcal{M}^{(-)(2)}_{\geq 2} - \left.\mathcal{M}^{(-)(2)}_{\geq 2}\right|_{s\rightarrow e^{2 \pi i}}\right].\nonumber
\end{align}
Since $\gamma_{(1,1)}$, $\mathcal{M}^{(-)(1)}_1$, and the two-loop cut are all real, the only imaginary parts of this formula come from the iterative $\gamma_{(1,1)}^2$ term and the difference term.  In particular, the difference term must be entirely imaginary, as at this order, only $J_{\kappa(2)}$, $S_{(2,2)}$, and $\bar{J}_{\kappa'(2)}$ can generate a $\log(s)$, the coefficient of which is purely imaginary. Then by looking at the real and imaginary parts separately, we obtain the following set of formulas:
\begin{align}
    &\mathcal{M}^{(-)(0)}_{1}\gamma_{(1,1)}^{(2)} +\mathcal{M}^{(-)(1)}_{1}\gamma_{(1,1)}^{(1)} = \frac{1}{\pi}\left(\mathcal{M}^\dag\mathcal{M}^{(-)}\right)^{(2)},\label{2 Loop Regge Re}\\
    &\left(\gamma_{(1,1)}^{(1)}\right)^2\mathcal{M}^{(-)(0)}_{1} = -\frac{2}{\pi^2}\left[\mathcal{M}^{(-)(2)}_{\geq 2} - \left.\mathcal{M}^{(-)(2)}_{\geq 2}\right|_{s\rightarrow e^{2 \pi i}}\right].\label{2 Loop Regge Im}
\end{align}
The reality of the term in the brackets follows from the fact that
the Glauber loop gives an $i\pi$ as
does the difference between the logs
in the two amplitudes.

We first use Eq. (\ref{2 Loop Regge Re}) to obtain the two loop anomalous dimension starting with  the diagrams needed for the one-loop single Glauber term $\mathcal{M}_1^{(-)(1)}$, of which there are two soft contributions:
\beq
\l(\begin{gathered}
\scalebox{0.5}{
\begin{tikzpicture}
\begin{feynman}
	\vertex (f1) at (0,1.5);
	\vertex (f2) at (0,-1.5);
	\vertex (p3) at (1, 1.5);
	\vertex (p2) at (-1, 1.5);
	\vertex (p1) at (-1, -1.5);
	\vertex (p4) at (1, -1.5);
	\vertex [label= \(n\)] at (-1, 1.);
	\vertex [label= \(\bar{n}\)] at (-1, -1.5);
	\vertex (g1) at (0, 1.278);
	\vertex (g2) at (0, -.722);
	\vertex (c1) at (.25, 1);
	\vertex (c2) at (.25, -1);
	\vertex (s1) at (0,.65);
	\vertex (s2) at (0, -.65);
\diagram*{
	(f1)--[scalar, red, line width = 0.3mm](s1),
	(f2)--[scalar, red, line width = 0.3mm](s2),
	(p2)--[charged scalar,  line width = 0.3mm](f1)--[charged scalar,  line width = 0.3mm](p3),
	(p4)--[charged scalar,  line width = 0.3mm](f2)--[charged scalar,  line width = 0.3mm](p1),
	(s2)--[SGreen, fermion,looseness = 1.75, line width=0.3mm, half right](s1)--[SGreen, fermion,looseness = 1.75, line width=0.3mm, half right](s2),
};
\end{feynman}
	\filldraw[red] (0,1.5) ellipse (0.6mm and 1.2mm);
	\filldraw[red] (0, -1.5) ellipse (0.6mm and 1.2mm);
	\filldraw[red] (0,.65) ellipse (0.6mm and 1.2mm);
	\filldraw[red] (0, -.65) ellipse (0.6mm and 1.2mm);
\end{tikzpicture}
}
\end{gathered}
\r)^{(-)}
= -i\mathcal{M}^{(-)(0)}\frac{2\,T_F n_f\alpha_s}{\pi} \,\frac{\Gamma(2-\epsilon)^2}{\Gamma(4-2\epsilon)}\Gamma(\epsilon)\l(\frac{\bar{\mu}^2}{-t}\r)^\epsilon.
\eeq
The soft eye graph is given by
\begin{align}
\l(\begin{gathered}
\scalebox{0.5}{
\begin{tikzpicture}
\begin{feynman}
	\vertex (f1) at (0,1.5);
	\vertex (f2) at (0,-1.5);
	\vertex (p3) at (1, 1.5);
	\vertex (p2) at (-1, 1.5);
	\vertex (p1) at (-1, -1.5);
	\vertex (p4) at (1, -1.5);
	\vertex [label= \(n\)] at (-1, 1.);
	\vertex [label= \(\bar{n}\)] at (-1, -1.5);
	\vertex (g1) at (0, 1.278);
	\vertex (g2) at (0, -.722);
	\vertex (c1) at (.25, 1);
	\vertex (c2) at (.25, -1);
	\vertex (s1) at (0,.65);
	\vertex (s2) at (0, -.65);
\diagram*{
	(f1)--[scalar, red, line width = 0.3mm](s1),
	(f2)--[scalar, red, line width = 0.3mm](s2),
	(p2)--[charged scalar,  line width = 0.3mm](f1)--[charged scalar,  line width = 0.3mm](p3),
	(p4)--[charged scalar,  line width = 0.3mm](f2)--[charged scalar,  line width = 0.3mm](p1),
	(s2)--[SGreen, gluon,looseness = 1.75, line width=0.3mm, half right](s1)--[SGreen, gluon,looseness = 1.75, line width=0.3mm, half right](s2),
};
\end{feynman}
	\filldraw[red] (0,1.5) ellipse (0.6mm and 1.2mm);
	\filldraw[red] (0, -1.5) ellipse (0.6mm and 1.2mm);
	\filldraw[red] (0,.65) ellipse (0.6mm and 1.2mm);
	\filldraw[red] (0, -.65) ellipse (0.6mm and 1.2mm);
\end{tikzpicture}
}
\end{gathered}\r)^{(-)}
=& -i\mathcal{M}^{(-)(0)}\frac{C_A \alpha_s}{2\pi}\l(\frac{\bar{\mu}^2}{-t}\r)^\epsilon\bigg\{\frac{\Gamma(2-\epsilon)^2}{\Gamma(4-2\epsilon)}\Gamma(\epsilon) - 2\frac{\Gamma(1-\epsilon)^2}{\Gamma(2-2\epsilon)}\Gamma(\epsilon) \nonumber\\
& + \frac{\Gamma(\eta/2)\Gamma(1/2-\eta/2)\Gamma(1 + \epsilon + \eta/2)\Gamma(-\epsilon-\eta/2)}{\Gamma(1 + \eta/2)\Gamma(1/2 -\epsilon-\eta/2) 4^{-\epsilon}}\l(\frac{\nu^2}{-t}\r)^{\eta/2}\bigg\}.
\label{S_1 sum}
\end{align}
Adding these together, we find
\begin{align}
&\l(
\begin{gathered}
\scalebox{0.5}{
\begin{tikzpicture}
\begin{feynman}
	\vertex (f1) at (0,1.5);
	\vertex (f2) at (0,-1.5);
	\vertex (p3) at (1, 1.5);
	\vertex (p2) at (-1, 1.5);
	\vertex (p1) at (-1, -1.5);
	\vertex (p4) at (1, -1.5);
	\vertex [label= \(n\)] at (-1, 1.);
	\vertex [label= \(\bar{n}\)] at (-1, -1.5);
	\vertex (g1) at (0, 1.278);
	\vertex (g2) at (0, -.722);
	\vertex (c1) at (.25, 1);
	\vertex (c2) at (.25, -1);
	\vertex (s1) at (0,.65);
	\vertex (s2) at (0, -.65);
\diagram*{
	(f1)--[scalar, red, line width = 0.3mm](s1),
	(f2)--[scalar, red, line width = 0.3mm](s2),
	(p2)--[charged scalar,  line width = 0.3mm](f1)--[charged scalar,  line width = 0.3mm](p3),
	(p4)--[charged scalar,  line width = 0.3mm](f2)--[charged scalar,  line width = 0.3mm](p1),
	(s2)--[SGreen, fermion,looseness = 1.75, line width=0.3mm, half right](s1)--[SGreen, fermion,looseness = 1.75, line width=0.3mm, half right](s2),
};
\end{feynman}
	\filldraw[red] (0,1.5) ellipse (0.6mm and 1.2mm);
	\filldraw[red] (0, -1.5) ellipse (0.6mm and 1.2mm);
	\filldraw[red] (0,.65) ellipse (0.6mm and 1.2mm);
	\filldraw[red] (0, -.65) ellipse (0.6mm and 1.2mm);
\end{tikzpicture}
}
\end{gathered}
\,\,+\,\,
\begin{gathered}
\scalebox{0.5}{
\begin{tikzpicture}
\begin{feynman}
	\vertex (f1) at (0,1.5);
	\vertex (f2) at (0,-1.5);
	\vertex (p3) at (1, 1.5);
	\vertex (p2) at (-1, 1.5);
	\vertex (p1) at (-1, -1.5);
	\vertex (p4) at (1, -1.5);
	\vertex [label= \(n\)] at (-1, 1.);
	\vertex [label= \(\bar{n}\)] at (-1, -1.5);
	\vertex (g1) at (0, 1.278);
	\vertex (g2) at (0, -.722);
	\vertex (c1) at (.25, 1);
	\vertex (c2) at (.25, -1);
	\vertex (s1) at (0,.65);
	\vertex (s2) at (0, -.65);
\diagram*{
	(f1)--[scalar, red, line width = 0.3mm](s1),
	(f2)--[scalar, red, line width = 0.3mm](s2),
	(p2)--[charged scalar,  line width = 0.3mm](f1)--[charged scalar,  line width = 0.3mm](p3),
	(p4)--[charged scalar,  line width = 0.3mm](f2)--[charged scalar,  line width = 0.3mm](p1),
	(s2)--[SGreen, gluon,looseness = 1.75, line width=0.3mm, half right](s1)--[SGreen, gluon,looseness = 1.75, line width=0.3mm, half right](s2),
};
\end{feynman}
	\filldraw[red] (0,1.5) ellipse (0.6mm and 1.2mm);
	\filldraw[red] (0, -1.5) ellipse (0.6mm and 1.2mm);
	\filldraw[red] (0,.65) ellipse (0.6mm and 1.2mm);
	\filldraw[red] (0, -.65) ellipse (0.6mm and 1.2mm);
\end{tikzpicture}
}
\end{gathered}
\r)^{(-)}\nonumber\\
 =&i\mathcal{M}^{(-)(0)}\,\l(\frac{\alpha_s}{4\pi}\r)\l(\frac{\mu^2}{-t}\r)^\epsilon\bigg[-2C_A\frac{\Gamma(1 + \epsilon)\Gamma(-\epsilon)^2}{\Gamma(-2\epsilon)}\l(\frac{1}{\eta} + \frac12 L_\nu\r)   \nonumber\\
&- \frac{2 C_A}{\epsilon^2} + \frac{11 C_A}{3\epsilon}-\frac{ 4T_F n_f}{3\epsilon} + \frac{67C_A }{9}- \zeta(2)C_A-\frac{20 T_F n_f}{9} \label{Soft Amp 2 Loop}\\
&+\epsilon\bigg( - \frac{28 C_A}{3}\zeta(3)-\frac{11 C_A}{6}\zeta(2) + \frac{404 C_A}{27} + \frac{2 T_F n_f}{3}\zeta(2) -\frac{112n_f T_F}{27} \bigg) + O(\epsilon^2)\bigg],\nonumber
\end{align}
where we have introduced $L_\nu = \log (\nu^2/-t)$.  

We have only three classes of cut diagrams to compute.  There are two diagrams involving the soft eye graph on either side of the cut.  These are just the soft eye graph convoluted with the Glauber box diagram, and they evaluate to
\begin{align}
&
\l(
\begin{gathered}
\scalebox{0.5}{
\begin{tikzpicture}
\begin{feynman}
	\vertex (f1) at (1,1.5);
	\vertex (f2) at (1,-1.5);
	\vertex (p3) at (1.75, 1.5);
	\vertex (p2) at (-1.75, 1.5);
	\vertex (p1) at (-1.75, -1.5);
	\vertex (p4) at (1.75, -1.5);
	\vertex [label= \(n\)] at (-1.75, 1);
	\vertex [label= \(\bar{n}\)] at (-1.75, -1.5);
	\vertex (g1) at (-1, 1.5);
	\vertex (g2) at (-1, -1.5);
	\vertex (c1) at (0, 2.0);
	\vertex (c2) at (0, -2.0);
 	\vertex (c3) at (0, 1.5);
	\vertex (c4) at (0, -1.5);
	\vertex (s1) at (-1,.622);
	\vertex (s2) at (-1, -.622);
\diagram*{
	(f1)--[scalar, red, line width = 0.3mm](f2),
	(p2)--[charged scalar,  line width = 0.3mm](g1)--[charged scalar,  line width = 0.3mm](c3)--[charged scalar,  line width = 0.3mm](f1)--[charged scalar,  line width = 0.3mm](p3),
	(p4)--[charged scalar,  line width = 0.3mm](f2)--[charged scalar,  line width = 0.3mm](c4)--[charged scalar,  line width = 0.3mm](g2)--[charged scalar,  line width = 0.3mm](p1),
	(c1)--[scalar, gray](c2),
	(g1)--[scalar, red, line width = 0.3mm](s1),
	(g2)--[scalar, red, line width = 0.3mm](s2),
	(s2)--[SGreen, gluon,looseness = 1.75, line width=0.3mm, half right](s1)--[SGreen, gluon,looseness = 1.75, line width=0.3mm, half right](s2),
};
\end{feynman}
	\filldraw[red] (-1,1.5) ellipse (0.6mm and 1.2mm);
	\filldraw[red] (-1,-1.5) ellipse (0.6mm and 1.2mm);
	\filldraw[red] (1,1.5) ellipse (0.6mm and 1.2mm);
	\filldraw[red] (1, -1.5) ellipse (0.6mm and 1.2mm);
	\filldraw[red] (-1,.622) ellipse (0.6mm and 1.2mm);
	\filldraw[red] (-1,-.622) ellipse (0.6mm and 1.2mm);
\end{tikzpicture}
}
\end{gathered}
\,\,+\,\,
\begin{gathered}
\scalebox{0.5}{
\begin{tikzpicture}
\begin{feynman}
	\vertex (f1) at (1,1.5);
	\vertex (f2) at (1,-1.5);
	\vertex (p3) at (1.75, 1.5);
	\vertex (p2) at (-1.75, 1.5);
	\vertex (p1) at (-1.75, -1.5);
	\vertex (p4) at (1.75, -1.5);
	\vertex [label= \(n\)] at (-1.75, 1);
	\vertex [label= \(\bar{n}\)] at (-1.75, -1.5);
	\vertex (g1) at (-1, 1.5);
	\vertex (g2) at (-1, -1.5);
	\vertex (c1) at (0, 2.0);
	\vertex (c2) at (0, -2.0);
	\vertex (s1) at (1,.622);
	\vertex (s2) at (1, -.622);
  	\vertex (c3) at (0, 1.5);
	\vertex (c4) at (0, -1.5);
\diagram*{
	(f1)--[scalar, red, line width = 0.3mm](s1),
	(f2)--[scalar, red, line width = 0.3mm](s2),
	(p2)--[charged scalar,  line width = 0.3mm](g1)--[charged scalar,  line width = 0.3mm](c3)--[charged scalar,  line width = 0.3mm](f1)--[charged scalar,  line width = 0.3mm](p3),
	(p4)--[charged scalar,  line width = 0.3mm](f2)--[charged scalar,  line width = 0.3mm](c4)--[charged scalar,  line width = 0.3mm](g2)--[charged scalar,  line width = 0.3mm](p1),
	(c1)--[scalar, gray](c2),
	(g1)--[scalar, red, line width = 0.3mm](g2),
	(s2)--[SGreen, gluon,looseness = 1.75, line width=0.3mm, half right](s1)--[SGreen, gluon,looseness = 1.75, line width=0.3mm, half right](s2),
};
\end{feynman}
	\filldraw[red] (-1,1.5) ellipse (0.6mm and 1.2mm);
	\filldraw[red] (-1,-1.5) ellipse (0.6mm and 1.2mm);
	\filldraw[red] (1,1.5) ellipse (0.6mm and 1.2mm);
	\filldraw[red] (1, -1.5) ellipse (0.6mm and 1.2mm);
	\filldraw[red] (1,.622) ellipse (0.6mm and 1.2mm);
	\filldraw[red] (1,-.622) ellipse (0.6mm and 1.2mm);
\end{tikzpicture}
}
\end{gathered}
\r)^{(-)}
\nonumber\\
&= -    \mathcal{M}^{(-)(0)}\frac{ C^2_A \alpha_s^2}{ 4\pi}\l(\frac{\bar{\mu}^2}{-t}\r)^{2\epsilon}\bigg\{\l(\frac{\Gamma(2-\epsilon)^2}{\Gamma(4-2\epsilon)}\Gamma(\epsilon) - 2\frac{\Gamma(1-\epsilon)^2}{\Gamma(2-2\epsilon)}\Gamma(\epsilon)\r)B(1,1+\epsilon) \\
& + \frac{\Gamma(\eta/2)\Gamma(1/2-\eta/2)\Gamma(1 + \epsilon + \eta/2)\Gamma(-\epsilon-\eta/2)}{\Gamma(1 + \eta/2)\Gamma(1/2 -\epsilon-\eta/2) 4^{-\epsilon}} B(1,1 + \epsilon + \eta/2)\l(\frac{\nu^2}{-t}\r)^{\eta/2}\bigg\}.\nonumber
\end{align}
Similarly, there are 2 cut diagrams with soft fermion loops:
\begin{align}
&\l(
\begin{gathered}
\scalebox{0.5}{
\begin{tikzpicture}
\begin{feynman}
	\vertex (f1) at (1,1.5);
	\vertex (f2) at (1,-1.5);
	\vertex (p3) at (1.75, 1.5);
	\vertex (p2) at (-1.75, 1.5);
	\vertex (p1) at (-1.75, -1.5);
	\vertex (p4) at (1.75, -1.5);
	\vertex [label= \(n\)] at (-1.75, 1);
	\vertex [label= \(\bar{n}\)] at (-1.75, -1.5);
	\vertex (g1) at (-1, 1.5);
	\vertex (g2) at (-1, -1.5);
	\vertex (c1) at (0, 2.0);
	\vertex (c2) at (0, -2.0);
	\vertex (s1) at (-1,.622);
	\vertex (s2) at (-1, -.622);
   	\vertex (c3) at (0, 1.5);
	\vertex (c4) at (0, -1.5);
\diagram*{
	(f1)--[scalar, red, line width = 0.3mm](f2),
	(p2)--[charged scalar,  line width = 0.3mm](g1)--[charged scalar,  line width = 0.3mm](c3)--[charged scalar,  line width = 0.3mm](f1)--[charged scalar,  line width = 0.3mm](p3),
	(p4)--[charged scalar,  line width = 0.3mm](f2)--[charged scalar,  line width = 0.3mm](c4)--[charged scalar,  line width = 0.3mm](g2)--[charged scalar,  line width = 0.3mm](p1),
	(c1)--[scalar, gray](c2),
	(g1)--[scalar, red, line width = 0.3mm](s1),
	(g2)--[scalar, red, line width = 0.3mm](s2),
		(s2)--[SGreen, fermion,looseness = 1.75, line width=0.3mm, half right](s1)--[SGreen, fermion,looseness = 1.75, line width=0.3mm, half right](s2),
};
\end{feynman}
	\filldraw[red] (-1,1.5) ellipse (0.6mm and 1.2mm);
	\filldraw[red] (-1,-1.5) ellipse (0.6mm and 1.2mm);
	\filldraw[red] (1,1.5) ellipse (0.6mm and 1.2mm);
	\filldraw[red] (1., -1.5) ellipse (0.6mm and 1.2mm);
	\filldraw[red] (-1,.622) ellipse (0.6mm and 1.2mm);
	\filldraw[red] (-1,-.622) ellipse (0.6mm and 1.2mm);
\end{tikzpicture}
}
\end{gathered}
\,\,+\,\,
\begin{gathered}
\scalebox{0.5}{
\begin{tikzpicture}
\begin{feynman}
	\vertex (f1) at (1,1.5);
	\vertex (f2) at (1,-1.5);
	\vertex (p3) at (1.75, 1.5);
	\vertex (p2) at (-1.75, 1.5);
	\vertex (p1) at (-1.75, -1.5);
	\vertex (p4) at (1.75, -1.5);
	\vertex [label= \(n\)] at (-1.75, 1);
	\vertex [label= \(\bar{n}\)] at (-1.75, -1.5);
	\vertex (g1) at (-1, 1.5);
	\vertex (g2) at (-1, -1.5);
	\vertex (c1) at (0, 2.0);
	\vertex (c2) at (0, -2.0);
	\vertex (s1) at (1,.622);
	\vertex (s2) at (1, -.622);
    \vertex (c3) at (0, 1.5);
	\vertex (c4) at (0, -1.5);
\diagram*{
	(f1)--[scalar, red, line width = 0.3mm](s1),
	(f2)--[scalar, red, line width = 0.3mm](s2),
	(p2)--[charged scalar,  line width = 0.3mm](g1)--[charged scalar,  line width = 0.3mm](c3)--[charged scalar,  line width = 0.3mm](f1)--[charged scalar,  line width = 0.3mm](p3),
	(p4)--[charged scalar,  line width = 0.3mm](f2)--[charged scalar,  line width = 0.3mm](c4)--[charged scalar,  line width = 0.3mm](g2)--[charged scalar,  line width = 0.3mm](p1),
	(c1)--[scalar, gray](c2),
	(g1)--[scalar, red, line width = 0.3mm](g2),
		(s2)--[SGreen, fermion,looseness = 1.75, line width=0.3mm, half right](s1)--[SGreen, fermion,looseness = 1.75, line width=0.3mm, half right](s2),
};
\end{feynman}
	\filldraw[red] (-1,1.5) ellipse (0.6mm and 1.2mm);
	\filldraw[red] (-1,-1.5) ellipse (0.6mm and 1.2mm);
	\filldraw[red] (1,1.5) ellipse (0.6mm and 1.2mm);
	\filldraw[red] (1, -1.5) ellipse (0.6mm and 1.2mm);
	\filldraw[red] (1,.622) ellipse (0.6mm and 1.2mm);
	\filldraw[red] (1,-.622) ellipse (0.6mm and 1.2mm);
\end{tikzpicture}
}
\end{gathered}\r)^{(-)}
=-\mathcal{M}^{(-)(0)} \frac{C_A\,T_F n_f\alpha^2_s}{\pi}\frac{\Gamma(2-\epsilon)^2}{\Gamma(4-2\epsilon)}\Gamma(\epsilon) B(1, 1 +\epsilon)\l(\frac{\bar{\mu}^2}{-t}\r)^{2\epsilon}.
\end{align}
Here, $B(a,b)$ is the coefficient of the one-loop bubble integral in $2-2\epsilon$ dimensions:
\beq
(4\pi)^{-\epsilon}\int\frac{[d^{2-\epsilon}k_\perp]}{[\vec{k}_\perp^2]^a [(\vec{k}_\perp + \vec{q}_\perp)^2]^b} = \frac{B(a,b)}{4\pi}(\vec{q}_\perp^2)^{1-\epsilon-a-b},
\label{Bubble Def}
\eeq
with 
\beq
B(a,b) = \frac{\Gamma(1-a-\epsilon)\Gamma(1-b-\epsilon)\Gamma(-1+a+b+\epsilon)}{\Gamma(a)\Gamma(b)\Gamma(2-a-b-2\epsilon)}.
\label{Bubble Coef}
\eeq

Lastly, we move onto the cut H-graph.  The uncut H-graph was computed in \cite{Moult2022AnomalousConstants}, and the cut graph is almost identical.  We have
\begin{align}
\l(
\begin{gathered}
\scalebox{0.5}{
\begin{tikzpicture}
\begin{feynman}
	\vertex (f1) at (1,1.5);
	\vertex (f2) at (1,-1.5);
	\vertex (p3) at (1.75, 1.5);
	\vertex (p2) at (-1.75, 1.5);
	\vertex (p1) at (-1.75, -1.5);
	\vertex (p4) at (1.75, -1.5);
	\vertex [label= \(n\)] at (-1.75, 1.);
	\vertex [label= \(\bar{n}\)] at (-1.75, -1.5);
	\vertex (g1) at (-1, 1.5);
	\vertex (g2) at (-1, -1.5);
	\vertex (c1) at (0, 2.0);
	\vertex (c2) at (0, -2.0);
	\vertex (s1) at (1,0);
	\vertex (s2) at (-1, 0);
    \vertex (c3) at (0, 1.5);
	\vertex (c4) at (0, -1.5);
\diagram*{
	(f1)--[scalar, red, line width = 0.3mm](s1),
	(f2)--[scalar, red, line width = 0.3mm](s1),
	(p2)--[charged scalar,  line width = 0.3mm](g1)--[charged scalar,  line width = 0.3mm](c3)--[charged scalar,  line width = 0.3mm](f1)--[charged scalar,  line width = 0.3mm](p3),
	(p4)--[charged scalar,  line width = 0.3mm](f2)--[charged scalar,  line width = 0.3mm](c4)--[charged scalar,  line width = 0.3mm](g2)--[charged scalar,  line width = 0.3mm](p1),
	(c1)--[scalar, gray](c2),
	(g1)--[scalar, red, line width = 0.3mm](g2),
	(s2)--[SGreen, gluon,line width=0.3mm](s1),
};
\end{feynman}
	\filldraw[red] (-1,1.5) ellipse (0.6mm and 1.2mm);
	\filldraw[red] (-1,-1.5) ellipse (0.6mm and 1.2mm);
	\filldraw[red] (1,1.5) ellipse (0.6mm and 1.2mm);
	\filldraw[red] (1, -1.5) ellipse (0.6mm and 1.2mm);
	\filldraw[red] (1,0) ellipse (0.6mm and 1.2mm);
	\filldraw[red] (-1,0) ellipse (0.6mm and 1.2mm);
\end{tikzpicture}
}
\end{gathered}\r)^{(-)}
 =&-\mathcal{M}^{(-)(0)}\frac{C_A^2\alpha_s^2}{16\pi}\frac{\Gamma(\eta/2)\Gamma(1/2-\eta/2)}{\sqrt{\pi}2^\eta}\l(\frac{\bar{\mu}^2}{-t}\r)^{2\epsilon}\l(\frac{\nu^2}{-t}\r)^{\eta/2}\\
&\times\l[B_2(\eta/2)e^{-2\gamma_E\epsilon}-2B(1,1 )B(1+ \eta/2 , 1  +\epsilon)\r].\nonumber
\end{align}
 $B_2$ is the coefficient of a two-loop bubble integral:
\begin{equation}
e^{2\gamma_E\epsilon}\int\frac{[d^{d^\prime} k_{1\perp}][d^{d^\prime}k_{2\perp}]}{\vec{k}_{1\perp}^2\vec{k}_{2\perp}^2(\vec{k}_{1\perp}-\vec{q}_\perp)^2(\vec{k}_{2\perp}-\vec{q}_\perp)^2[(\vec{k}_{1\perp} - \vec{k}_{2\perp})^2]^{\eta/2}} = \frac{B_2(\eta/2)}{(4\pi)^{2-2\epsilon}}(\vec{q}_\perp^2)^{-2-2e-\eta/2}.
\end{equation}
To linear order in $\eta$, $B_2$ is given by \cite{Kazakov:1983pk, Kotikov:2018wxe}
\begin{align}
B_2\l(\frac{\eta}{2}\r) &= B(1,1)^2\,e^{2\gamma_E\epsilon} + \frac{\eta}{2}\bigg(B(1,1)^2\,e^{2\gamma_E\epsilon}\l(\psi^{(0)}(1  +2\epsilon) -\psi^{(0)}(-\epsilon)\r)\nonumber\\
& + \frac{\Gamma(-\epsilon)^2\Gamma(-1-\epsilon)\Gamma(2\epsilon)}{\Gamma(-1-3\epsilon)}\,_3F_2\l(1,1,-2\epsilon;1-2\epsilon, 2 + \epsilon; 1\r)\bigg) + O(\eta^2),\\
&=B(1,1)^2\,e^{2\gamma_E\epsilon} +\frac{\eta}{2}\l(-\frac{1}{\epsilon^3} + \frac{\zeta(2)}{\epsilon} -\frac{76\zeta(3)}{3}+O(\epsilon)\r) + O(\eta^2).\nonumber
\end{align}
Adding up the cut graphs, we then have
\begin{align}
\text{Cut graphs} =&\mathcal{M}^{(-)(0)}\l(\frac{\pi}{2}\r) \l(\frac{\alpha_s}{4\pi}\r)^2\l(\frac{\mu^2}{-t}\r)^{2\epsilon}\bigg[-\l(\frac{1}{\eta}  +\frac12\log\frac{\nu^2}{-t}\r)\l(2\alpha^{(1)}\r)^2 +\frac{8 C_A^2}{\epsilon^3}- \frac{22 C_A^2}{\epsilon^2} \nonumber\\
& +\frac{8C_A\, T_F n_f}{\epsilon^2} -\frac{134 C_A^2}{3\epsilon} + \frac{4C_A^2 \zeta(2)}{\epsilon} + \frac{40 C_A\, T_f n_f }{3\epsilon} -\frac{808C_A^2}{9}-22C_A^2\zeta(2)\label{Cut 2 Loop}\\
& + \frac{68C_A^2}{3}\zeta(3)+\frac{224 C_A\,T_F n_f}{9} - 8C_A\,T_F n_f \zeta(2)\bigg].\nonumber
\label{S_2 sum}
\end{align}
where $\alpha_1=- \gamma_{(1,1)}\frac{4\pi}{\alpha}$.   The cut graphs are rapidity divergent only because we have not included the collinear
contributions. Nonetheless the Eq.(\ref{Master Regge Odd}) assures us that the LHS of Eq.(\ref{2 Loop Regge Re}) will also be rapidity divergent.  
  From the two loop formula in Eq. (\ref{2 Loop Regge Re}), we then find that the two-loop RAD to be given by
\begin{align}
\gamma_{(1,1)}^{(2)} &= \l(\frac{\alpha_s}{4\pi}\r)^2C_A\l(\frac{\mu^2}{-t}\r)^{2\epsilon}\bigg[-\frac{11C_A}{3\epsilon^2} + \frac{4n_f T_F}{3\epsilon^2} - \frac{67C_A}{9\epsilon} +\frac{2C_A}{\epsilon}\zeta(2)+\frac{20T_F n_f}{9\epsilon}\nonumber\\
&  - \frac{404 C_A}{27}+\frac{11C_A}{3}\zeta(2)+2C_A\zeta(3)+\frac{112 T_F n_f}{27} - \frac{4T_F n_f}{3}\zeta(2)\bigg].
\end{align}
This agrees with the standard result \cite{Fadin:1995xg, Fadin:1996tb,Fadin:1995km, Blumlein:1998ib}, after accounting for different choices of normalization.  As a further check, we can expand our results to $O(\epsilon^2)$ and compare with the results given in \cite{Falcioni:2021dgr}, where we again find  agreement.
\subsection{The Role of the Collinear Modes}

Notice that when we calculated the cut diagrams we did not include collinear contributions we will not show dont contribute to $\gamma_{(1,1)}^{(2)}$.  In \cite{Moult2022AnomalousConstants,Gao2024} it was argued that in the planar limit, all collinear functions $J_{\kappa(i)}$ must reduce to something proportional to ${\cal M}_1$. That is, there is no convolution
and the result must lead to a Regge pole. 
Intuitively this follows from the collapse rule which tells us that Glaubers come in a burst and can only be connected to one collinear line.
As such, the contribution to the collinear function from planar graphs only depends upon the physical transverse momentum exchanges $q_\perp$ and serves to reproduce the iterative term $\mathcal{M}^{(-)}_1\gamma_{(1,1)}^{(1)}$  in Eq.(\ref{2 Loop Regge Re}).
  We now show how this comes about by way of an example.  Consider the planar double Glauber vertex correction graph:
\begin{align}
\l(\begin{gathered}
\scalebox{.5}{
\begin{tikzpicture}
\begin{feynman}
	\vertex [label = below: \(n\)] (i1)  at (-1.75,1.1);
	\vertex  [label = below: \(\)] (f1) at (1.75, 1.1);
	\vertex [label = above: \(\bar{n}\)] (i2)  at (-1.75,-1.5);
	\vertex [label = above: \(\)]  (f2) at (1.75, -1.5);
	\node[dot] (cn1) at (-1.25,1.1);
	\node[dot] (cn2) at (1.25,1.1);
	\vertex (g11) at (-.75,1.1);
	\vertex (g12) at (.75,1.1);
	\vertex (g21) at (-.75,-1.5);
	\vertex (g22) at (.75,-1.5);
	\vertex (c1) at (0, 1.75);
	\vertex (c2) at (0,-1.75);
    \vertex (c3) at (0, 1.1);
	\vertex (c4) at (0, -1.5);
\diagram*{
	(i1)--[scalar, line width = 0.3mm](g11)--[charged scalar, line width = 0.3mm](c3)--[charged scalar, line width = 0.3mm](g12)--[scalar, line width = 0.3mm](f1),
	(f2)--[charged scalar, line width = 0.3mm](g22)--[charged scalar, line width = 0.3mm](c4)--[charged scalar, line width = 0.3mm](g21)--[charged scalar, line width = 0.3mm](i2),
	(g11)--[scalar, red, line width = 0.3mm](g21),
	(g12)--[scalar, red, line width = 0.3mm](g22),
	(c1)--[scalar, gray](c2),
(cn2)--[line width = 0.3mm, quarter right, looseness = 1](cn1),
(cn2)--[gluon, line width = 0.3mm, quarter right, looseness = 1](cn1),
};
\end{feynman}
	\filldraw[red] (-.75,1.1) ellipse (0.6mm and 1.2mm);
	\filldraw[red] (-.75,-1.5) ellipse (0.6mm and 1.2mm);
	\filldraw[red] (.75,1.1) ellipse (0.6mm and 1.2mm);
	\filldraw[red] (.75, -1.5) ellipse (0.6mm and 1.2mm);
\end{tikzpicture}
}
\end{gathered}\r)^{(-)}
=& \mathcal{M}_1^{(-)(0)}\frac{C_A}{4}(8\pi \alpha_s)  \int\dbar^d k\,\dbar^d \ell\frac{\mathcal{N}(\ell)}{\vec{k}_\perp^2(\vec{k}_\perp + \vec{q}_\perp)^2(\ell + p)^2(\ell + p + q)^2}\nonumber\\
&\times (2\pi)^3 \delta_+(\ell^2) \,\bar{n}\cdotp( p + \ell)\,\delta_+\l((p + k + \ell)^2\r) \,n\cdotp p'\,\delta_+\l((p' + k)^2\r).
\end{align}
Here $\ell$ is the collinear loop momentum, and $k$ is the Glauber loop momentum.  $\mathcal{N}(\ell)$ is the numerator, and it is independent of $k$ regardless of whether the collinear projectiles are gluons or quarks.  Therefore the only $k$-dependence in the integrand is in the Glauber propagators and the delta functions, and we can integrate over $k$ to obtain
\begin{align}
&
\l(\begin{gathered}
\scalebox{.5}{
\begin{tikzpicture}
\begin{feynman}
	\vertex [label = below: \(n\)] (i1)  at (-1.75,1.1);
	\vertex  [label = below: \(\)] (f1) at (1.75, 1.1);
	\vertex [label = above: \(\bar{n}\)] (i2)  at (-1.75,-1.5);
	\vertex [label = above: \(\)]  (f2) at (1.75, -1.5);
	\node[dot] (cn1) at (-1.25,1.1);
	\node[dot] (cn2) at (1.25,1.1);
	\vertex (g11) at (-.75,1.1);
	\vertex (g12) at (.75,1.1);
	\vertex (g21) at (-.75,-1.5);
	\vertex (g22) at (.75,-1.5);
	\vertex (c1) at (0, 1.75);
	\vertex (c2) at (0,-1.75);
    \vertex (c3) at (0, 1.1);
	\vertex (c4) at (0, -1.5);
\diagram*{
	(i1)--[scalar, line width = 0.3mm](g11)--[charged scalar, line width = 0.3mm](c3)--[charged scalar, line width = 0.3mm](g12)--[scalar, line width = 0.3mm](f1),
	(f2)--[charged scalar, line width = 0.3mm](g22)--[charged scalar, line width = 0.3mm](c4)--[charged scalar, line width = 0.3mm](g21)--[charged scalar, line width = 0.3mm](i2),
	(g11)--[scalar, red, line width = 0.3mm](g21),
	(g12)--[scalar, red, line width = 0.3mm](g22),
	(c1)--[scalar, gray](c2),
(cn2)--[line width = 0.3mm, quarter right, looseness = 1](cn1),
(cn2)--[gluon, line width = 0.3mm, quarter right, looseness = 1](cn1),
};
\end{feynman}
	\filldraw[red] (-.75,1.1) ellipse (0.6mm and 1.2mm);
	\filldraw[red] (-.75,-1.5) ellipse (0.6mm and 1.2mm);
	\filldraw[red] (.75,1.1) ellipse (0.6mm and 1.2mm);
	\filldraw[red] (.75, -1.5) ellipse (0.6mm and 1.2mm);
\end{tikzpicture}
}
\end{gathered}\r)^{(-)}\nonumber\\
&= \mathcal{M}_1^{(-)(0)} \frac{C_A}{4}(8\pi \alpha_s)  \frac{B(1,1)}{2(4\pi)}\l(\frac{\bar{\mu}^2}{-t}\r)^\epsilon \int\dbar^d \ell\,\frac{\mathcal{N}(\ell)\,\theta(-\bar{n}\cdotp \ell)\,(2\pi) \delta_+(\ell^2)}{((\ell + p)^2(\ell + p + q)^2},\label{Collinear Glauber Reduced}\\
&=\l[\pi\gamma_{(1,1)}^{(1)}\r]\,\l( \mathcal{M}_1^{(-)(0)} \int\dbar^d \ell\,\frac{\mathcal{N}(\ell)\,\theta(-\bar{n}\cdotp \ell)\,(2\pi) \delta_+(\ell^2)}{(\ell + p)^2(\ell + p + q)^2}\r).\nonumber
\end{align}
 If we now turn to the one-loop collinear contribution to $\mathcal{M}_{(1)}^{(-)}$, we find
\beq
\l(\begin{gathered}
\scalebox{.5}{
\begin{tikzpicture}
\begin{feynman}
	\vertex [label = below: \(n\)] (i1)  at (-1.75,1.1);
	\vertex  [label = below: \(\)] (f1) at (1.75, 1.1);
	\vertex [label = above: \(\bar{n}\)] (i2)  at (-1.75,-1.5);
	\vertex [label = above: \(\)]  (f2) at (1.75, -1.5);
	\node[dot] (cn1) at (-1.25,1.1);
	\node[dot] (cn2) at (1.25,1.1);
	\vertex (g1) at (0,1.1);
	\vertex (g2) at (0,-1.5);
	\vertex (c1) at (0, 1.75);
	\vertex (c2) at (0,-1.75);
\diagram*{
	(i1)--[charged scalar, line width = 0.3mm](g1)--[charged scalar, line width = 0.3mm](f1),
	(f2)--[charged scalar, line width = 0.3mm](g2)--[charged scalar, line width = 0.3mm](i2),
	(g1)--[scalar, red, line width = 0.3mm](g2),
(cn2)--[line width = 0.3mm, quarter right, looseness = 1](cn1),
(cn2)--[gluon, line width = 0.3mm, quarter right, looseness = 1](cn1),
};
\end{feynman}
	\filldraw[red] (0,1.1) ellipse (0.6mm and 1.2mm);
	\filldraw[red] (0, -1.5) ellipse (0.6mm and 1.2mm);
\end{tikzpicture}
}
\end{gathered}\r)^{(-)}
= \mathcal{M}_1^{(-)(0)}(-i)\int\dbar^d \ell\,\frac{\mathcal{N}(\ell)}{(\ell + p)^2(\ell + p + q)^2\ell^2}.
\label{Collinear Glauber One Loop}
\eeq
This is almost identical to the term in the round brackets in Eq. (\ref{Collinear Glauber Reduced}), and indeed they are equal.  We can see this by doing the contour integration over $n\cdotp \ell$ in Eq. (\ref{Collinear Glauber One Loop})
 Thus we have
\beq
\l(\begin{gathered}
\scalebox{.5}{
\begin{tikzpicture}
\begin{feynman}
	\vertex [label = below: \(n\)] (i1)  at (-1.75,1.1);
	\vertex  [label = below: \(\)] (f1) at (1.75, 1.1);
	\vertex [label = above: \(\bar{n}\)] (i2)  at (-1.75,-1.5);
	\vertex [label = above: \(\)]  (f2) at (1.75, -1.5);
	\node[dot] (cn1) at (-1.25,1.1);
	\node[dot] (cn2) at (1.25,1.1);
	\vertex (g11) at (-.75,1.1);
	\vertex (g12) at (.75,1.1);
	\vertex (g21) at (-.75,-1.5);
	\vertex (g22) at (.75,-1.5);
	\vertex (c1) at (0, 1.75);
	\vertex (c2) at (0,-1.75);
    \vertex (c3) at (0, 1.1);
	\vertex (c4) at (0, -1.5);
\diagram*{
	(i1)--[scalar, line width = 0.3mm](g11)--[charged scalar, line width = 0.3mm](c3)--[charged scalar, line width = 0.3mm](g12)--[scalar, line width = 0.3mm](f1),
	(f2)--[charged scalar, line width = 0.3mm](g22)--[charged scalar, line width = 0.3mm](c4)--[charged scalar, line width = 0.3mm](g21)--[charged scalar, line width = 0.3mm](i2),
	(g11)--[scalar, red, line width = 0.3mm](g21),
	(g12)--[scalar, red, line width = 0.3mm](g22),
	(c1)--[scalar, gray](c2),
(cn2)--[line width = 0.3mm, quarter right, looseness = 1](cn1),
(cn2)--[gluon, line width = 0.3mm, quarter right, looseness = 1](cn1),
};
\end{feynman}
	\filldraw[red] (-.75,1.1) ellipse (0.6mm and 1.2mm);
	\filldraw[red] (-.75,-1.5) ellipse (0.6mm and 1.2mm);
	\filldraw[red] (.75,1.1) ellipse (0.6mm and 1.2mm);
	\filldraw[red] (.75, -1.5) ellipse (0.6mm and 1.2mm);
\end{tikzpicture}
}
\end{gathered}\r)^{(-)}
= \l[(i\pi)\gamma_{(1,1)}^{(1)}\r]\,\times \,
\l(\begin{gathered}
\scalebox{.5}{
\begin{tikzpicture}
\begin{feynman}
	\vertex [label = below: \(n\)] (i1)  at (-1.75,1.1);
	\vertex  [label = below: \(\)] (f1) at (1.75, 1.1);
	\vertex [label = above: \(\bar{n}\)] (i2)  at (-1.75,-1.5);
	\vertex [label = above: \(\)]  (f2) at (1.75, -1.5);
	\node[dot] (cn1) at (-1.25,1.1);
	\node[dot] (cn2) at (1.25,1.1);
	\vertex (g1) at (0,1.1);
	\vertex (g2) at (0,-1.5);
	\vertex (c1) at (0, 1.75);
	\vertex (c2) at (0,-1.75);
\diagram*{
	(i1)--[charged scalar, line width = 0.3mm](g1)--[charged scalar, line width = 0.3mm](f1),
	(i2)--[charged scalar, line width = 0.3mm](g2)--[charged scalar, line width = 0.3mm](f2),
	(g1)--[scalar, red, line width = 0.3mm](g2),
(cn2)--[line width = 0.3mm, quarter right, looseness = 1](cn1),
(cn2)--[gluon, line width = 0.3mm, quarter right, looseness = 1](cn1),
};
\end{feynman}
	\filldraw[red] (0,1.1) ellipse (0.6mm and 1.2mm);
	\filldraw[red] (0, -1.5) ellipse (0.6mm and 1.2mm);
\end{tikzpicture}
}
\end{gathered}\r)^{(-)}
\eeq
which matches exactly with what we expect for the one-loop iterative terms. An identical set of manipulations work for the other collinear planar  graph.

 Let us consider the two-loop non-planar collinear cut graphs.  These must give a vanishing contribution if we wish to maintain
 a Regge pole behavior.

These have vanishing contributions to the odd-signature amplitude:
\begin{equation}
\l(\begin{gathered}
\scalebox{.5}{
\begin{tikzpicture}
\begin{feynman}
	\vertex [label = below: \(n\)] (i1)  at (-1.75,1.5);
	\vertex  [label = below: \(\)] (f1) at (1.75, 1.5);
	\vertex [label = above: \(\bar{n}\)] (i2)  at (-1.75,-1.5);
	\vertex [label = above: \(\)]  (f2) at (1.75, -1.5);
	\node[dot] (cn1) at (-1.25,1.5);
	\node[dot] (cn2) at (1.25,1.5);
	\vertex (m1) at (0, 1.5);
	\vertex (m2) at (0,-1.5);
	\vertex (g11) at (-.75,1.1);
	\vertex (g12) at (.75,1.5);
	\vertex (g21) at (-.75,-1.5);
	\vertex (g22) at (.75,-1.5);
	\vertex (c1) at (0, 1.75);
	\vertex (c2) at (0,-1.75);
    \vertex (c3) at (0, 1.5);
	\vertex (c4) at (0, -1.5);
\diagram*{
	(i1)--[charged scalar, line width = 0.3mm](g12)--[scalar, line width = 0.3mm](f1),
	(f2)--[charged scalar, line width = 0.3mm](g22)--[charged scalar, line width = 0.3mm](c4)--[charged scalar, line width = 0.3mm](g21)--[charged scalar, line width = 0.3mm](i2),
	(g11)--[scalar, red, line width = 0.3mm](g21),
	(g12)--[scalar, red, line width = 0.3mm](g22),
	(c1)--[scalar, gray](c2),
(cn1)--[line width = 0.3mm, quarter right, looseness = 1](cn2),
(cn1)--[gluon, line width = 0.3mm, quarter right, looseness = 1](cn2),
};
\end{feynman}
	\filldraw[red] (-.75,1.1) ellipse (0.6mm and 1.2mm);
	\filldraw[red] (-.75,-1.5) ellipse (0.6mm and 1.2mm);
	\filldraw[red] (.75,1.5) ellipse (0.6mm and 1.2mm);
	\filldraw[red] (.75, -1.5) ellipse (0.6mm and 1.2mm);
\end{tikzpicture}
}
\end{gathered}
+
\begin{gathered}
\scalebox{.5}{
\begin{tikzpicture}
\begin{feynman}
	\vertex [label = below: \(n\)] (i1)  at (-1.75,1.5);
	\vertex  [label = below: \(\)] (f1) at (1.75, 1.5);
	\vertex [label = above: \(\bar{n}\)] (i2)  at (-1.75,-1.5);
	\vertex [label = above: \(\)]  (f2) at (1.75, -1.5);
	\node[dot] (cn1) at (-1.25,1.5);
	\node[dot] (cn2) at (1.25,1.5);
	\vertex (m1) at (0, 1.5);
	\vertex (m2) at (0,-1.5);
	\vertex (g11) at (-.75,1.5);
	\vertex (g12) at (.75,1.1);
	\vertex (g21) at (-.75,-1.5);
	\vertex (g22) at (.75,-1.5);
	\vertex (c1) at (0, 1.75);
	\vertex (c2) at (0,-1.75);
    \vertex (c3) at (0, 1.5);
	\vertex (c4) at (0, -1.5);
\diagram*{
	(i1)--[scalar, line width = 0.3mm](g11)--[charged scalar, line width = 0.3mm](f1),
	(f2)--[charged scalar, line width = 0.3mm](g22)--[charged scalar, line width = 0.3mm](c4)--[charged scalar, line width = 0.3mm](g21)--[charged scalar, line width = 0.3mm](i2),
	(g11)--[scalar, red, line width = 0.3mm](g21),
	(g12)--[scalar, red, line width = 0.3mm](g22),
	(c1)--[scalar, gray](c2),
(cn1)--[line width = 0.3mm, quarter right, looseness = 1](cn2),
(cn1)--[gluon, line width = 0.3mm, quarter right, looseness = 1](cn2),
};
\end{feynman}
	\filldraw[red] (-.75,1.5) ellipse (0.6mm and 1.2mm);
	\filldraw[red] (-.75,-1.5) ellipse (0.6mm and 1.2mm);
	\filldraw[red] (.75,1.1) ellipse (0.6mm and 1.2mm);
	\filldraw[red] (.75, -1.5) ellipse (0.6mm and 1.2mm);
\end{tikzpicture}
}
\end{gathered}\r)^{(-)}
=0.
\end{equation}


\section{Bootstrapping the Anomalous Dimensions}
\subsection{ Determining $\gamma_{(2,2)}^{(1)}$ from $\gamma_{(1,1)}^{(1)}$ in the octet channel}

Consider
the following form of the master formula at second order in the coupling
\begin{align}
\l(\gamma_{(1,1)}^{(1)}\r)^2\mathcal{M}_1^{(-)(0)} &= \frac{2}{\pi^2}\l[\l(e^{i\pi\bar{K}_z}-1\r)\mathcal{M}_{\geq 2}^{(-)}\r]^{(2)},\nonumber\\
&= \frac{2i}{\pi}\gamma_{(2,2)}^{(1)}\otimes\mathcal{M}_{\geq 2}^{(-)(0)},
\label{Iterative Relations}
\end{align}
where the second line is defined as
\begin{align}
\label{relation}
    \gamma_{(2,2)}^{(1)}\otimes\mathcal{M}_{\geq 2}^{(-)(0)} &=\frac{i}{2} J_{\kappa(2)}^{(0)(s)A_1 A_2}\otimes\l(\gamma_{(2,2)}^{A_1 A_2;B_1B_2}\otimes S^{(0)(s)B_1 B_2; D_1D_2}_{(2,2)}+S^{(0)(s)A_1 A_2 ;C_1C_2}_{(2,2)}\otimes \gamma_{(2,2)}^{C_1C_2;D_1D_2}\r)\nn \\ &\otimes\bar{J}_{\kappa'(2)}^{(0)(s)D_1 D_2}- (s\leftrightarrow u).
\end{align}
Here  we repristinated the color indices as they will play in important role in our analysis.
Now we would like to ask what constraints does this equation put on $\gamma_{(2,2)}$? At this order both the soft function and the jet functions are trivially given by (\ref{treeJ}) and (\ref{treeS}) respectively.  In Eq. (\ref{Iterative Relations}) both side of the equation are projected down to the $8_A$  irrep.    due to the minus signature.

Thus  Eq. (\ref{Iterative Relations}) is given by
\beq
 \left(\frac{\bar \mu^2}{-t}\right)^{2 \epsilon}\frac{(C_A\alpha_s )^2}{(4\pi)^2}B[1,1]^2{\cal M}_1^{(-)(0)} =g^4\frac{2}{\pi}\frac{C_A}{2}\mu^{4 \epsilon}{\cal M}_0^{(-)}\frac{ ( q_\perp^2)}{16 \pi \alpha_s }
\int \frac{[d^{d^\prime}l_\perp]}{l_\perp^2(l_\perp-q_\perp)^2}\frac{[d^{d^\prime}k_\perp]}{k_\perp^2(k_\perp-q_\perp)^2}\gamma^{8}_{(2,2)}(k_\perp,l_\perp,q_\perp),
 \eeq

We immediately notice a remarkable simplification. The LHS is proportional to $B[1,1]^2$, which implies that the integral projects out the piece of $\gamma_{(2,2)}^{8_A}$ which is independent of $k_\perp$ and $l_\perp$, since any non-trivial dependence on these variables would necessarily, by dimensional analysis, lead to  terms proportional to $B[1,1+\epsilon]$.
So we are left with the relation
\bea
\label{final}
 \left(\frac{\bar \mu^2}{q_\perp^2}\right)^{2 \epsilon}\frac{(\alpha_a C_A)^2}{(4\pi)^2}B[1,1]^2 {\cal M}_0^{(-)}
 &=&\frac{g^4}{\pi}C_A\mu^{4 \epsilon}\frac{ ( q_\perp^2)}{16 \pi\alpha_s }
\int \frac{[d^{d^\prime}l_\perp]}{l_\perp^2(l_\perp-q_\perp)^2}\frac{[d^{d^\prime}k_\perp]}{k_\perp^2(k_\perp-q_\perp)^2}\gamma^{8}_{(2,2)}(q_\perp){\cal M}_0^{(-)} \nn \\
&=& \frac{g^4}{\pi}C_A\mu^{4 \epsilon}\frac{ ( q_\perp^2)}{16 \pi \alpha_s}((4\pi)^{-1+\epsilon}(q_\perp^{-2-2\epsilon}B[1,1])^2
\gamma^{8}_{(2,2)}(q_\perp){\cal M}_0^{(-)} \nn \\
 \eea
Therefore
\beq
\label{res}
\gamma^{8 P}_{(2,2)}= C_A\alpha_s   q_\perp^2
\eeq
where the $P$ reminds us  that in principle there are other terms in $\gamma^{8 }_{(2,2)}$ whose integral yields zero in Eq. (\ref{final}).
Note that the fact that the convolutive piece gets projected out implies that there is a power law (Regge pole) solution at NLL in 
accordance with Eq.(\ref{Fadin}). We can now see how the running up to NLL reproduces the form of $(\ref{Fadin})$.
The RRG equation for $J^{A_1 A_2 8_A}$ is given by
\bea
\nu \frac{d}{d\nu} J^{A_1 A_2 8_A}(k_\perp)&=& \int \frac{[d^{d^\prime}k_\perp]}{k_\perp^2(k_\perp-q_\perp)^2} \gamma_{(2,2)}(k_\perp,q_\perp) J^{A_1 A_2 8_A}(k_\perp) \nn \\
&=&\gamma_{(1,1)} ~ J^{A_1 A_2 8_A},
\eea
which had to be the case if Eq.(\ref{Fadin}) is to hold.

If we want to predict the full $\gamma_{(2,2)}$ we need some additional data, which can be gleaned by recalling that
that $J^{A_1 A_28_A}$ can be factorized into a time ordered product of two $O^{qB}_n$ operators defined in Eq.(\ref{eq:On})
\beq
J^{A_1 A_28_A}(k_\perp)=\int dx^\pm_1 dx^\pm_2 \langle p \mid T( O^{qA_1}_n(k_\perp,x_1^\pm) O^{qA_2}_n(k_\perp-q_\perp,x_2^\pm) \mid p^\prime \rangle^{8_A},
\eeq
This operator will run due to the anomalous dimensions of the individual $O^{qA_1}_n$, as well as an additional renormalization due to
the semi-local nature of the operator. The key distinction between the two types of renormalizations is that the former will not
generate any convolution while the ladder will. The anomalous dimensions of $O^{qA_1}_n$ were calculated in \cite{Rothstein:2016bsq},
\beq
\gamma_{O^{q }_n}=\gamma_{(1,1)}
\eeq
which is as we would expect for one Glauber exchange. We will generate  such contribution to the anomalous dimensions
from each of the Glaubers, the first of which carries momentum $k_\perp$ and the other $k_{\perp}-q_\perp$.
These contributions are shown in the first two diagrams of figure (3).

\begin{figure}
\begin{subfigure}[b]{.31\textwidth}
\caption{\qquad \qquad \qquad \qquad}
\centering
\scalebox{1}{
\begin{tikzpicture}
\begin{feynman}
	\vertex (i1)  at (-2,1.5);
	\vertex (f1) at (2, 1.5);
	\vertex (i2)  at (-2,-1.5);
	\vertex (f2) at (2, -1.5);
	\vertex (g11) at (-1,1.5);
	\vertex (g12) at (1,1.5);
	\vertex (g21) at (-1,-1.5);
	\vertex (g22) at (1,-1.5);
	\vertex (m1) at (0, 1.5);
	\vertex (m2) at (0,-1.5);
	\vertex (c1) at (0, 2.125);
	\vertex (c2) at (0,-2.125);
	\vertex [label= \(n\)] at (-2, 1);
	\vertex [label= \(\bar{n}\)] at (-2, -1.5);
	\vertex (k2)  at (-1,1.5);
	\vertex (k1) at (1,1.5);
	\vertex[label = \(\gamma_{(2,2)}^{T} \)] at (0,.65);
\diagram*{
	(i1)--[charged scalar, line width = 0.3mm](g11),
    (g12)--[charged scalar, line width = 0.3mm](f1),
	(f2)--[charged scalar, line width = 0.3mm](g22)--[charged scalar, line width = 0.3mm](g21)--[charged scalar, line width = 0.3mm](i2),
    (g11)--[scalar, red, line width = 0.3mm,momentum'={[arrow style=black,arrow shorten=0.35] \tiny\(k_\perp\)}](g21),
   (g22)--[scalar, red, line width = 0.3mm,momentum'={[arrow style=black,arrow shorten=0.35]\tiny \(k_\perp + q_\perp\)}](g12),
	(k2)--[double, purple, line width = .3mm](k1)
};
\end{feynman}
	\filldraw[red] (-1,1.5) ellipse (0.6mm and 1.2mm);
	\filldraw[red] (-1,-1.5) ellipse (0.6mm and 1.2mm);
	\filldraw[red] (1,1.5) ellipse (0.6mm and 1.2mm);
	\filldraw[red] (1, -1.5) ellipse (0.6mm and 1.2mm);
	\filldraw[purple] (-1,1.5) ellipse (1.2mm and 2.4mm);
	\filldraw[purple] (1,1.5) ellipse (1.2mm and 2.4mm);
\end{tikzpicture}
}
\end{subfigure}
\begin{subfigure}[b]{.31\textwidth}
\caption{\qquad \qquad \qquad\qquad }
\centering
\scalebox{1}{
\begin{tikzpicture}
\begin{feynman}
	\vertex (i1)  at (-2,1.5);
	\vertex (f1) at (2, 1.5);
	\vertex (i2)  at (-2,-1.5);
	\vertex (f2) at (2, -1.5);
	\vertex (g11) at (-1,1.5);
	\vertex (g12) at (1,1.5);
	\vertex (g21) at (-1,-1.5);
	\vertex (g22) at (1,-1.5);
	\vertex (m1) at (0, 1.5);
	\vertex (m2) at (0,-1.5);
	\vertex (c1) at (0, 2.125);
	\vertex (c2) at (0,-2.125);
	\vertex [label= \(n\)] at (-2, 1);
	\vertex [label= \(\bar{n}\)] at (-2, -1.5);
	\vertex [label = \(\omega_G(k_\perp)\)] at (-.2,.75);
\diagram*{
	(i1)--[charged scalar, line width = 0.3mm](g11)--[charged scalar, line width = 0.3mm](g12)--[charged scalar, line width = 0.3mm](f1),
	(f2)--[charged scalar, line width = 0.3mm](g22)--[charged scalar, line width = 0.3mm](g21)--[charged scalar, line width = 0.3mm](i2),
	(g11)--[scalar, red, line width = 0.3mm,momentum'={[arrow style=black,arrow shorten=0.35] \tiny\(k_\perp\)}](g21),
	(g22)--[scalar, red, line width = 0.3mm,momentum'={[arrow style=black,arrow shorten=0.35]\tiny \(k_\perp + q_\perp\)}](g12),
};
\end{feynman}
	\filldraw[red] (-1,1.5) ellipse (0.6mm and 1.2mm);
	\filldraw[red] (-1,-1.5) ellipse (0.6mm and 1.2mm);
	\filldraw[red] (1,1.5) ellipse (0.6mm and 1.2mm);
	\filldraw[red] (1, -1.5) ellipse (0.6mm and 1.2mm);
	\filldraw[blue] (-1,1.5) ellipse (1.5mm and 3.0mm);
\end{tikzpicture}
}
\end{subfigure}
\begin{subfigure}[b]{.31\textwidth}
\caption{\qquad \qquad \qquad \qquad}
\centering
\scalebox{1}{
\begin{tikzpicture}
\begin{feynman}
	\vertex (i1)  at (-2,1.5);
	\vertex (f1) at (2, 1.5);
	\vertex (i2)  at (-2,-1.5);
	\vertex (f2) at (2, -1.5);
	\vertex (g11) at (-1,1.5);
	\vertex (g12) at (1,1.5);
	\vertex (g21) at (-1,-1.5);
	\vertex (g22) at (1,-1.5);
	\vertex (m1) at (0, 1.5);
	\vertex (m2) at (0,-1.5);
	\vertex (c1) at (0, 2.125);
	\vertex (c2) at (0,-2.125);
	\vertex [label= \(n\)] at (-2, 1);
	\vertex [label= \(\bar{n}\)] at (-2, -1.5);
	\vertex [label = \(\omega_G(k_\perp + q_\perp)\)] at (2.4,.75);
\diagram*{
	(i1)--[charged scalar, line width = 0.3mm](g11)--[charged scalar, line width = 0.3mm](g12)--[charged scalar, line width = 0.3mm](f1),
	(f2)--[charged scalar, line width = 0.3mm](g22)--[charged scalar, line width = 0.3mm](g21)--[charged scalar, line width = 0.3mm](i2),
	(g11)--[scalar, red, line width = 0.3mm,momentum'={[arrow style=black,arrow shorten=0.35] \tiny\(k_\perp\)}](g21),
	(g22)--[scalar, red, line width = 0.3mm,momentum'={[arrow style=black,arrow shorten=0.35]\tiny \(k_\perp + q_\perp\)}](g12),
};
\end{feynman}
	\filldraw[red] (-1,1.5) ellipse (0.6mm and 1.2mm);
	\filldraw[red] (-1,-1.5) ellipse (0.6mm and 1.2mm);
	\filldraw[red] (1,1.5) ellipse (0.6mm and 1.2mm);
	\filldraw[red] (1, -1.5) ellipse (0.6mm and 1.2mm);
	\filldraw[blue] (1,1.5) ellipse (1.5mm and 3.0mm);
\end{tikzpicture}
}
\end{subfigure} 
\caption{}
\end{figure}

In addition there is another contribution to $\gamma_{(2,2)}$ stemming from diagrams that are of the form
of the last diagram in figure (3) which span both exchanges and lead to a non-trivial convolution.
Thus we make the following ansatz for the two Glauber jet RRG equation 
\bea
\nu \frac{d}{d\nu} J^{A_1 A_2 8_A}(l_\perp)&=&\int [d^{2-2\epsilon}k] \l[\delta^{2-2\epsilon}\l(k_\perp-l_\perp)(\gamma_{(1,1)}(k_\perp)+\gamma_{(1,1)}(k_\perp-q_\perp)\r)+K(l_\perp,k_\perp)\r]J^{A_1 A_2 8_A}(k_\perp). \nn \\
\eea
Now we can fix $K(l_\perp,k_\perp)$ by imposing that Eq.(\ref{final}) be obeyed and find
\beq
K(l_\perp,k_\perp)=\alpha_s C_A \l[ -\frac{l_\perp^2}{k_\perp^2(l_\perp-k_\perp)^2}- \frac{(q_\perp-l_\perp)^2}{k_\perp^2(k_\perp-q_\perp+l_\perp)^2}+ \frac{q_\perp^2}{k_\perp^2(q_\perp-k_\perp)^2}\r].
\eeq
which  agrees with the explicit calculation performed in  \cite{Gao2024}. We have thus shown that we can fix the full two Glauber octet anomalous dimension
directly from the one Glauber operator. That this had to happen for the integrated anomalous dimension assuming the result in Eq.(\ref{Fadin}), but that the full result could follow from the one Glauber operator is a new result
as far as we are aware.

\subsection{Constraining Other Color Channels from Positive Signature Amplitudes}

Let us now explore the information stored in the positive signature amplitude relation Eq.(\ref{Master Regge Even})
\begin{equation}
\label{real}
    \l(e^{i\pi \bar{K}_z}-1\r)\mathcal{M}^{(+)} = -2\text{Re}[\mathcal{M}^{(+)}].
\end{equation}
We garner new information from this result  
since the minus signature is not sensitive
to channels other than the octet at the two Glauber level.
  The plus signature amplitude starts at two Glauber exchange   at order $\alpha^2$ however this contribution if purely imaginary. Thus the RHS of Eq.(\ref{Master Regge Eve}).
starts at $O(\alpha^3$) i.e. two loops.
\beq
\label{plus}
\frac{4i}{ \pi}\Re[\mathcal{M}]^{(2)(+)}=
((J^{(0)}_{\kappa(2)}\otimes \gamma^{(1)(1,8_S)}_{(2,2)}) \otimes S^{(0)}_{(2,2)}\otimes\bar{J}_{\kappa'(2)}^{(0)}+
(J^{(0)}_{\kappa(i)} \otimes S^{(0)}_{(2,2)}\otimes(\gamma^{(1)(1,8_S)}_{(2,2)}\otimes \bar{J}_{\kappa'(2)}^{(0)}))+
(s \rightarrow u),
\eeq

Only the $8_S$ and  singlet representation   in the decomposition $8 \otimes 8$ contribute for quark states.
Extracting the LHS from the full theory calculations \cite{Caola:2022dfa,Glover:2004si,Ahmed:2019qtg}, and using the known results for  $\gamma_{(2,2)}^{8_S}$ (a Regge pole) and singlet (Regge cut)\cite{Ioffe:2010zz} leads to  agreement with Eq.(\ref{plus}). By calculating the amplitude up to order $\epsilon^2$, we can use the same strategy as in the case of the $8_A$ to determine the singlet and $8_S$ as well.

Notice that this method of calculation
  does not arise   from
a cut diagram and thus lacks that calculational advantages of the negative signature case. But it does
lead to significant simplification relative to the canonical method
of calculating the anomalous dimensions from the $1/\eta$ pole in that
there is no need for a rapidity regulator since one can simply use the
full theory result. Of course the full theory result involves many more diagrams than the EFT which can isolate the sources of the rapidity divergence.

We may generate other sum rules at higher order as well.
At three loops $O(\alpha^4)$
 the RRGE involves mixing between different $\mathcal{M}_{(i,j)}$,
so to facilitate this we introduce the  notation
\begin{align}
\gamma_{(i,j)}\otimes\mathcal{M}_{(i,j)}^{(+)} &= \frac12 J^{(s)}_{\kappa(i )} \otimes \left(S_{(i,k)}\otimes\gamma_{(k,j)} + \gamma_{(i,k)}\otimes S_{(k,j)}\right)\otimes\bar{J}_{\kappa'(j)}^{(s)} + (s\leftrightarrow u).\nonumber
\end{align}
 Then the two loop even signature formula at NLO is given by 
\begin{align}
 -2\Re[\mathcal{M}]^{(3)} = &i\pi\bigg[\gamma_{(2,2)}^{(2)}\otimes \mathcal{M}_{(2,2)}^{(+)(0)}  + \gamma_{(3,3)}^{(1)}\otimes \mathcal{M}_{(3,3)}^{(0)}+ \gamma_{(2,3)}^{(1)}\otimes \mathcal{M}^{(+)(0)}_{(2,3)}\nonumber\\
&     + \gamma_{(2,2)}^{(1)}\otimes \mathcal{M}_{(2,2)}^{(+)(1)}  + \frac{i\pi}{2!} \gamma_{(2,2)}^{(1)}\otimes \gamma_{(2,2)}^{(1)}\otimes \mathcal{M}_{(2,2)}^{(+)(0)} \bigg].
\end{align}
We can generate two constraints by considering the  real and imaginary parts since the anomalous dimension are purely real and the Glauber loops are are purely imaginary, 
\begin{align}
&\gamma_{(2,2)}^{(2)}\otimes \mathcal{M}_{(2,2)}^{(+)(0)} + \gamma_{(2,2)}^{(1)}\otimes \mathcal{M}_{(2,2)}^{(+)(1)} =  \frac{2i}{\pi}\Re[\mathcal{M}]^{(3)}\label{Even Two Loop Real}\\
&\gamma_{(3,3)}^{(1)}\otimes \mathcal{M}_{(3,3)}^{(0)}+ \gamma_{(2,3)}^{(1)}\otimes \mathcal{M}^{(+)(0)}_{(2,3)}  + \frac{i\pi}{2!} \gamma_{(2,2)}^{(1)}\otimes \gamma_{(2,2)}^{(1)}\otimes \mathcal{M}_{(2,2)}^{(+)(0)} =0.\label{Even Two Loop Imaginary}
\end{align}
Note that ${\cal M}_{(2,3)}$ is purely real as it contains two Glauber loops.
Here both formulas provide non-trivial relations about the various anomalous dimensions.  In particular, Eq. (\ref{Even Two Loop Imaginary}) provides an interesting relation between the one-loop anomalous dimensions that appear at this order, while Eq. (\ref{Even Two Loop Real}) contains the information for the two loop $\gamma_{(2,2)}$.  Notably, Eq. (\ref{Even Two Loop Imaginary}) constrains terms with different numbers of convolution integrals.  However, unlike the lower order relation, in this case there are many color channels (multiple singlets and octets) which will mix.

\section{Discussion}

In this paper, we have uncovered further structure in the two to two scattering amplitude in the  Regge limit.
This has been accomplished by working with the EFT for near forward scattering in a double expansion in $\alpha$ and $t/s$ which allows us always work with well defined gauge invariant matrix elements.
We have demonstrated that anomalous dimensions can be extracted via cut diagrams thus effectively reducing the loop order.  Our result (\ref{Master Regge Odd}) generalizes the result in \cite{Moult2022AnomalousConstants} in that it
works at all orders and for all color channels.
We have also shown that the two Glauber rapidity anomalous dimensions can be bootstrapped from the one Glauber case. These results follow from a combination of factorization, unitarity and crossing symmetry.
We have given generalized relations between amplitude and anomalous dimension for both the positive and negative signature amplitudes, each of which generates a separate set of constraints on the anomalous dimension. The positive signature constraint allowed us to simply extract the $8_A$ channel two Glauber anomalous dimensions and
our result agree with those in \cite{Ioffe:2010zz, Gao2024}.


It would be interesting to use this formalism to compute the three loop Regge trajectory. Progress in this direction has been made within the context of Reggeon field theory \cite{Falcioni:2021dgr}.
It seems reasonable to believe that our results genearted in the context of an EFT of QCD  could also yield insights into the Reggeon approach. 
In the EFT approach using our formalism should help significantly reduce the complexity of the calculation, reducing most of the work to computing $\mathcal{M}^{(-)}_1$ through two loops.  In fact, the full three loop QCD amplitude is known, and so in principle one could avoid computing the cut by simply extracting the imaginary part.  

Another avenue worth pursuing is to explore the even signature formula.  In principle, one could use this for a very clean derivation of the leading order BFKL equation, and as a useful probe of the mixing between different Glauber operators at next to leading order and beyond.  At three loops, one has matrix elements that are both imaginary, such as $S_{(2,2)}$, and real, such as $S_{(3,3)}$, and so one may be able to find interesting relations between anomalous dimensions that appear at different orders through iterative terms.  

\section{Acknowledgements}
The authors would like to thanks Iain Stewart for discussions. The figures were made using \cite{Ellis:2016jkw}. The authors are supported by supported by the Department of Energy (Grant No.~DE-FG02-04ER41338 and FG02-06ER41449).

\bibliography{references.bib}

\end{document}